\documentclass[aps,prb,reprint,superscriptaddress,floatfix]{revtex4-1}
\usepackage[german,english]{babel}
\usepackage[utf8]{inputenc}
\usepackage{amsmath,amssymb}
\usepackage{graphicx}

\let\vec=\mathbf

\newcommand{\eps}{\eta^+}
\newcommand{\abs}[1]{\left| #1 \right|}
\newcommand{\mean}[1]{\left\langle #1 \right\rangle}

\newcommand{\floor}[1]{\left\lfloor #1 \right\rfloor}

\bibpunct{[}{]}{,}{n}{}{}

\begin{document}
\title{Impurity-induced subgap states in superconductors with inhomogeneous pairing}

\author{A. A. Bespalov}

\affiliation{Institute for Physics of Microstructures, Russian Academy of Sciences, 603950 Nizhny Novgorod, GSP-105, Russia}

\begin{abstract}
We study subgap states induced by a single impurity in an s-wave superconductor with suppressed pairing. For concreteness, we consider a bulk superconductor containing a normal spherical region. We find that a point impurity in this system induces two Yu-Shiba-Rusinov states inside the minigap instead of one, which one would have in a homogeneous superconductor. Moreover, the subgap states appear even if the impurity is nonmagnetic. We prove that this result actually holds almost for any superconductor with a real and spatially inhomogeneous order parameter, if the quasiparticle spectrum is gapped.

\end{abstract}

\maketitle

\section{Introduction}

Magnetic impurities are known to produce a pair-breaking effect in superconductors, as has been demonstrated by Abrikosov and Gor'kov in their seminal work \cite{AbrikosovGorkov61}. Later, a more subtle effect has been discovered: the existence of subgap states localized by magnetic impurities -- the so-called Yu-Shiba-Rusinov states \cite{Yu1965,Shiba68PTP,Rusinov1969JETP} (or Shiba states, in short). In recent years there has been an increased interest in systems hosting subgap Shiba states and bands. The interest stems from the realization that chains of magnetic atoms on superconductors can give rise to zero energy Majorana modes \cite{Nadj-Perge+2013PRB,Pientka+2013PRB,Vazifeh+PRL2013,Klinovaja+2013PRL,Braunecker+2013PRL} and indications of experimental observations of these modes \cite{Nadj-Perge+2014Science,Ruby+2015PRL,Feldman+2016Nature,Kim+2018Science} 
(see also Ref. \cite{Heinrich+PSS2018} for review of experimental progress in studying Shiba states). Majorana modes are considered as potential building blocks for topological quantum computers \cite{Chetan+2008RMP,Alicea2012RPP,Beenakker2013Review,Elliott+2015RMP}.

Another property of magnetic impurities, which is of practical significance, is their ability to trap nonequilibrium quasiparticles. Such quasiparticles are generated during operation of superconducting devices \cite{Court+2008PRB,Rajauria+2009PRB,Martinis+2009PRL,Barends+2011APL,Naruse+2012JLTP,Saira+2012PRB,ONeil+2012PRB,deVisser+2012JLTP,Riste+2013NatComm,Nguyen+2013NJP,Nsanzieneza+2014PRL,Wang+2014NatCommun} (including qubits, photon counters and coolers) and typically degrade their performance. Degradation appears mainly due to quasiparticles with energies close to or larger than the bulk gap $\Delta$, which can move freely in the superconductor. In a Josephson junction, for example, such quasiparticle can get stuck in a current-carrying Andreev state (``poison" it), which will change the current-phase characteristic of the junction. To avoid this, quasiparticle can be trapped far from the junction area. Magnetic impurities, providing Shiba states with energies lying well below the gap $\Delta$, may act as such traps. Indeed, a quasiparticle captured in a Shiba state is very unlikely to escape, if the temperature is much lower than $\Delta$. Eventually such quasiparticle will recombine with another one. Theoretical considerations show, in fact, that trapping results in an enhanced quasiparticle-quasiparticle recombination rate \cite{Kozorezov+2008PRB,Hijmaring+2009IEEE,Kozorezov+2009EPL}. Note also that free quasiparticles are the source of ohmic losses in superconductors, and trapping them in Shiba states will reduce the losses at low frequencies \cite{Fominov+PRB2011}. A paper by Barends et al. \cite{Barends+2009PRB} provides experimental evidence that magnetic disorder indeed can reduce the quasiparticle lifetimes in superconductors . Remarkably, nonmagnetic disorder seems to produce the same effect, which cannot be directly explained by the trapping mechanism, because nonmagnetic scatterers do not induce subgap states in a homogeneous superconductor \cite{Anderson1959JPCS}. 

It should be noted that the most commonly used type of quasiparticle trap is a normal drain in contact with the superconducting device \cite{Court+2008PRB,Rajauria+2009PRB,Saira+2012PRB,Nguyen+2013NJP}. Sufficiently large normal drains have deep subgap Andreev states, such that quasiparticles trapped in these states are unlikely to escape from the normal region (the same trapping principle can be implemented by suppressing superconductivity locally with a magnetic field \cite{Ullom+98APL,Nsanzieneza+2014PRL,Wang+2014NatCommun,Taupin+2016NatCommun}). The idea of combining two trapping mechanisms -- magnetic impurities and normal regions -- has not been previously considered in literature. In the present paper, we address this idea and find that in a superconductor the combination of a region with suppressed pairing and of an impurity (not necessarily magnetic) results in synergetic behavior, displaying features that are not present in each of these systems separately.

Specifically, we calculate the subgap electronic density of states in an s-wave superconductor with a normal inclusion in the shape of a ball (a normal bubble) and with a pointlike impurity. Similar systems with the impurity located in the center of the region with suppressed superconductivity have been studied theoretically before \cite{Flatte+PRL1997,Flatte+PRB1997,Salkola+PRB1997,Bespalov+2016}. Flatt\'e and Byers \cite{Flatte+PRL1997,Flatte+PRB1997} calculated numerically and self-consistently the local density of states in an s-wave superconductor with a finite-sized magnetic impurity. They found multiple impurity states whose energies are almost independent of the product $k_F \xi$ as long as $k_F \xi \gg 1$, where $k_F$ is the Fermi wavenumber and $\xi$ is the coherence length. This result agrees with the analytical calculations of Rusinov \cite{Rusinov1969JETP}, who found that in the limit $k_F \xi \gg 1$ the energy of the Shiba state with orbital momentum $l$ is given by
\begin{equation}
	E_{Sl} = \Delta_0 \abs{\cos(\alpha_{l\uparrow} - \alpha_{l\downarrow})},
	\label{eq:Rusinov}
\end{equation}
where $\Delta_0$ is the bulk gap, and $\alpha_{l\uparrow}$ and $\alpha_{l \downarrow}$ are the scattering phases of the impurity in the normal state for electrons with orbital momentum $l$ and spin up ($\uparrow$) or spin down ($\downarrow$), respectively. Note that none of the quantities in the right-hand side of Eq. \eqref{eq:Rusinov} depend on the coherence length. Thus, it appears that self-consistency has a very weak effect on Shiba states in typical s-wave superconductors. Remarkably, for a relatively low value of $k_F \xi = 10$ it has been found \cite{Flatte+PRB1997} that a \textit{nonmagnetic} impurity hosts a subgap state, which appears due to local suppression of the gap. However, the energy of this state is extremely close to the gap edge, so that this state is highly delocalized.

In Ref. \cite{Salkola+PRB1997}, a self-consistent calculation of impurity states in s-wave and d-wave superconductors has been performed within a tight-binding model. In Ref. \cite{Bespalov+2016}, Andreev states localized at a nonmagnetic impurity with a pairing amplitude $\Delta$ distorted on a scale $L_{\Delta} \ll \xi$ have been studied (note that self-consistent calculations typically yield $L_{\Delta} \sim k_F^{-1}$ and not $L_{\Delta} \sim \xi$ \cite{Flatte+PRL1997,Flatte+PRB1997,Schlottmann76}). It has been found that an impurity with a suppressed pairing amplitude $\Delta$ hosts an infinite number of subgap states (if the superconductor is unbounded), whose energies approach the bulk gap exponentially fast with growing orbital momentum $l$.

Contrary to previous works, here we consider a superconductor containing a spherical region with suppressed pairing which appears not due to an impurity, but has rather external causes, e.g. a normal inclusion or local heating. We assume that the superconducting gap is completely suppressed in a bubble with radius $a \gg k_F^{-1}$, with the most interesting case corresponding to $a \sim \xi$. The spectrum of such system without impurities has been studied in a number of papers \cite{Saint-James64,Hui+85PRB,Gunsenheimer+96,Bespalov+2016}. It has been found that the spectrum contains a large number (strictly speaking, an infinite number \cite{Bespalov+2016}) of subgap states. 
This situation may be reasonably well described within the quasiclassical approximation, where the orbital momentum $l$ varies continuously, such that the spectrum is continuous and has a minigap $E_g<\Delta$ \cite{Gunsenheimer+96}. In the present work, we add a point impurity to this system. According to Eq. \eqref{eq:Rusinov}, for a magnetic impurity in a homogeneous superconductor this yields one Shiba state, since only scattering with $l=0$ takes place. In the presence of a normal bubble one might expect one impurity state with energy $E<E_g$. This is indeed the case when the impurity is situated in the center of the bubble. However, when the impurity is shifted from the center and even put outside the bubble, we generally find two Shiba states with different energies inside the minigap. Moreover, these states are present even if the impurity is nonmagnetic -- then, these states have equal energies (due to spin degeneracy), which can be still well below $E_g$. This result means that a region with suppressed pairing may significantly enhance the ability of impurities to trap quasiparticles and even allows nonmagnetic scatterers to act as quasiparticle traps. At first sight the appearance of subgap states in the presence of a nonmagnetic impurity may seem inconsistent with Anderson's theorem \cite{Anderson1959JPCS}, however, this theorem does not apply to spatially inhomogeneous superconductors. We also want to point out that a closely related phenomenon has been previously found in one-dimensional SNS junctions: here, the energy of the lowest Andreev state may become even lower in the presence of a potential barrier \cite{Zaikin+80JETP,Bagwell92PRB}.

The specificity of the system that we consider may suggest that our results are mostly of academic significance, however, we show that they can be generalized for a broad class of system. In particular, we prove that a point impurity (no matter whether magnetic of not) in an s-wave superconductor with an inhomogeneous and real order parameter and with a gap in the quasiparticle spectrum almost always induces two localized subgap states.

The paper is organized as follows. In Sec. \ref{sec:Basic} we describe our technique (based on the Gor'kov equation) and provide a general expression for the Green function in the presence of a point impurity. In Sec. \ref{sec:pure} we review the known results concerning the spectrum of a normal bubble inside a superconductor without impurities. In Sec. \ref{sec:Impurity} we analyze the impurity-induced states, calculate their energies and wave function. In the conclusion the main results are summarized. The appendices contain most technical details of the calculations.

\section{Basic equations}
\label{sec:Basic}

The system that we study is an infinite s-wave superconductor containing a point impurity whose position is given by $\vec{r}_i$ and a region with suppressed pairing with radius $a \gg k_F^{-1}$. As an approximation, we use a steplike order parameter profile:
\begin{equation}
	\Delta(r) = \left\{
	\begin{array}{l}
	  0 \quad \mbox{when} \quad r \leq a, \\
		\Delta = \mathrm{const}>0 \quad \mbox{when} \quad r>a.
	\end{array} \right.
	\label{eq:Delta}
\end{equation}
Of course, such profile is not self-consistent and hence one should not expect accurate quantitative prediction from this model. However, we will prove that the main qualitative resuts captured by this model are quite general and hold even for profiles of $\Delta(\vec{r})$ that are not spherically symmetric.

We analyze the density of states in our system using the Green functions technique. The retarded Green function $\check{G}_E(\vec{r},\vec{r}')$ is determined from the Gor'kov equation \cite{Kopnin-book}:
\begin{eqnarray}
	& \left\{H_0(\vec{r}) + U(\vec{r}-\vec{r}_i) + \hat{\tau}_z [\vec{J}(\vec{r}-\vec{r}_i) \hat{\boldsymbol{\sigma}} - E - i\eps] \right. & \nonumber \\
	& \left. + \left(
	\begin{array}{cc}
	  0 & - \Delta(\vec{r}) \\
		\Delta^*(\vec{r}) & 0
	\end{array} \right) \right\} \check{G}_E(\vec{r},\vec{r}') = \delta(\vec{r} - \vec{r}'),&
	\label{eq:Gorkov}
\end{eqnarray}
\begin{equation}
	H_0(\vec{r}) = -\frac{\hbar^2 \nabla^2}{2m}  - \mu.
	\label{eq:H0_S}
\end{equation}
Here, $\hat{\tau}_z$ is a Pauli matrix in Nambu space, $U(\vec{r})$ is the electrical potential of the impurity, and $\vec{J}(\vec{r})$ is its exchange field, $\hat{\boldsymbol{\sigma}} = \{ \hat{\sigma}_x,\hat{\sigma}_y,\hat{\sigma}_z \}$ are the Pauli matrices in spin space, $E$ is the energy, $\eps$ is an infinitely small positive quantity, $m$ is the electron mass and $\mu = \hbar^2 k_F^2/(2m)$ is the chemical potential. The $4 \times 4$ Green function has the following block structure in Nambu space:
\begin{equation}
	\check{G}_E(\vec{r},\vec{r}') = \left(
	\begin{array}{cc}
	  \hat{G}_E(\vec{r},\vec{r}') & \hat{F}_E(\vec{r},\vec{r}') \\
		-\hat{F}_E^\dagger(\vec{r},\vec{r}') & \hat{\bar{G}}_E(\vec{r},\vec{r}')
	\end{array} \right).
	\label{eq:GE_blocks}
\end{equation}
Each block is a $2 \times 2$ matrix in spin space. We use the same definitions of the blocks in terms of electron field operators as in Ref. \cite{Bespalov2018}. 

The local density of states is given by
\begin{equation}
	\nu(E,\vec{r}) = \pi^{-1} \mathrm{Im} \left[ G_{E\uparrow\uparrow}(\vec{r},\vec{r}) + G_{E\downarrow\downarrow}(\vec{r},\vec{r}) \right],
	\label{eq:nu_def}
\end{equation}
where the arrows are the spin indices. %of $\hat{G}_E(\vec{r},\vec{r}')$.

Let us specify the properties of the point impurity. We assume that the range of the potentials $U(\vec{r})$ and $\vec{J}(\vec{r}) \hat{\boldsymbol{\sigma}}$ is much smaller than the electron wavelength. Then the impurity is essentially a spherical scatterer. For a nonmagnetic defect with $\vec{J}(\vec{r}) = 0$, in particular, this means that the solution of the scattering problem for a plane wave in vacuum has the form \cite{Landau3}
\begin{equation}
	\psi(\vec{r}) = e^{i\vec{k} \vec{r}} + \frac{e^{ikr + i\alpha} \sin \alpha}{kr}.
	\label{eq:plane_wave}
\end{equation}
Here, the defect is located at the origin, and $k$ and $\alpha \in [-\pi/2,\pi/2]$ are the energy dependent wavenumber and scattering phase, respectively. For a magnetic impurity one can choose the spin quantization axis in such a way that spin-up and spin-down electrons would be scattered without spin rotation and would have some scattering phases $\alpha_{\uparrow}$ and $\alpha_{\downarrow}$, respectively. With this choice of the spin quantization axis one can see that the equations for Green functions with spin indices $\uparrow \uparrow$ and $\downarrow \downarrow$ decouple, and the Green functions with indices $\uparrow \downarrow$ and $\downarrow \uparrow$ vanish. This greatly simplifies the solution of the Gor'kov equation.

To study the subgap spectrum of our system, we need to solve Eq. \eqref{eq:Gorkov} only for $\abs{E}<\Delta$. Typically, $\Delta \ll \mu$, so that it is reasonable to neglect the variations of $\alpha_{\uparrow}$ and $\alpha_{\downarrow}$ with energy for $\abs{E}<\Delta$. Within this approximation Eq. \eqref{eq:Gorkov} has been solved in Ref. \cite{Bespalov2018} in terms of the Green functions of the system without the impurity. In particular, it has been found that the function $G_{E\uparrow \uparrow}(\vec{r},\vec{r}')$ has the form
\begin{equation}
	G_{E\uparrow\uparrow}(\vec{r},\vec{r}') = G_E^{(0)}(\vec{r},\vec{r}') + G_{E\uparrow\uparrow}^{(1)}(\vec{r},\vec{r}'),
	\label{eq:GE_+impurity}
\end{equation}
where
\begin{widetext}
\begin{eqnarray}
	& G_{E\uparrow\uparrow}^{(1)}(\vec{r},\vec{r}') = {\cal D}_{\uparrow}^{-1}(E) \left( G_E^{(0)}(\vec{r},\vec{r}_i) \left\{ G_E^{(0)}(\vec{r}_i,\vec{r}') \left[ \frac{mk_F}{2\pi \hbar^2} \cot \alpha_{\downarrow} - G_{-ER}^{(0)*}(\vec{r}_i,\vec{r}_i) \right] - F_{-E}^{\dagger(0)*}(\vec{r}_i,\vec{r}_i) F_E^{\dagger(0)}(\vec{r}_i,\vec{r}') \right\} \right. & \nonumber \\
	& \left. - F_{-E}^{\dagger(0)*}(\vec{r},\vec{r}_i) \left\{ F_E^{\dagger(0)}(\vec{r}_i,\vec{r}') \left[ \frac{mk_F}{2\pi \hbar^2} \cot \alpha_{\uparrow} - G_{ER}^{(0)}(\vec{r}_i,\vec{r}_i) \right] + F_E^{\dagger(0)}(\vec{r}_i,\vec{r}_i) G_E^{\dagger(0)}(\vec{r}_i,\vec{r}') \right\} \right), &
	\label{eq:GE(1)}
\end{eqnarray}
\begin{equation}
	{\cal D}_{\uparrow}(E) = \left[ \frac{mk_F \cot \alpha_{\uparrow}}{2\pi \hbar^2} - G_{ER}^{(0)}(\vec{r}_i,\vec{r}_i) \right] \left[ \frac{mk_F \cot \alpha_{\downarrow}}{2\pi \hbar^2}
	- G_{-ER}^{(0)*}(\vec{r}_i,\vec{r}_i) \right] + F_E^{\dagger (0)} (\vec{r}_i,\vec{r}_i) F_{-E}^{\dagger (0)*} (\vec{r}_i,\vec{r}_i),
	\label{eq:D_up}
\end{equation}
\begin{equation}
	G_{ER}^{(0)}(\vec{r},\vec{r}') = G_E^{(0)}(\vec{r},\vec{r}') - \frac{m}{2\pi \hbar^2 \abs{\vec{r} - \vec{r}'}}.
	\label{eq:GER_def}
\end{equation}
The functions $G_E^{(0)}(\vec{r},\vec{r}')$ and $F_E^{\dagger(0)}(\vec{r},\vec{r}')$ solve the Gor'kov equation without the impurity:
\begin{equation}
	\left[ H_0(\vec{r}) - \hat{\tau}_z ( E + i\eps) + \left(
	\begin{array}{cc}
	  0 & - \Delta(\vec{r}) \\
		\Delta^*(\vec{r}) & 0
	\end{array} \right) \right] \left(
	\begin{array}{c}
	  G_E^{(0)}(\vec{r},\vec{r}') \\
		- F_E^{\dagger(0)}(\vec{r},\vec{r}')
	\end{array} \right)
	= \left(
	\begin{array}{c}
	  \delta(\vec{r}- \vec{r}') \\
		0
	\end{array} \right).
	\label{eq:Gorkov0}
\end{equation}
\end{widetext}
Thus, to calculate the density of states in the presence of an impurity, one should first find the Green functions of a pure system.

\section{Green functions and density of states in a pure normal bubble}
\label{sec:pure}

\subsection{Global subgap spectrum}
\label{sub:global}

The subgap spectrum of a normal bubble inside a superconductor has been studied using the Bogoliubov-de Gennes (BdG) equations formalism in several papers \cite{Saint-James64,Hui+85PRB,Gunsenheimer+96,Bespalov+2016}. Here, we will outline the main results. Clearly, the energy levels below the bulk gap are discrete. Due to the spherical symmetry, it is convenient to describe the spectrum by three quantum numbers: the orbital momentum $l$, its projection $l_z$ and a third number $n$. Since the Hamiltonian does not depend on $l_z$, each energy level is at least $2l+1$ times degenerate. Within the Andreev approximation, a relatively simple equation for the energy levels $E_{ln}$ has been obtained in Ref. \cite{Gunsenheimer+96}:
\begin{equation}
	E_{ln} = \frac{2 \Delta \xi'}{\pi L(l)} \left[ \pi n + \gamma(E_{ln}) \right],
	\label{eq:E_Gunsenheimer}
\end{equation}
where
\begin{equation}
	L(l) = 2\sqrt{a^2 - (l+1/2)^2 k_F^{-2}},
	\label{eq:L(l)}
\end{equation}
\begin{equation}
	\gamma(E) = \arccos \left( \frac{E}{\Delta} \right),
	\label{eq:gamma}
\end{equation}
$\xi' = \pi \hbar v_F/(2\Delta)$, $v_F = \hbar k_F/m$ is the Fermi velocity, $l = 0,1,2...$ and $n$ is any integer such that a real solution of Eq. \eqref{eq:E_Gunsenheimer} exists. There are also some restrictions on $l$, namely $k_F L(l) \gg 1$ and $[k_F L(l)]^3 \gg (l+ 1/2)^2$ \footnote{These restrictions are the applicability conditions for Debye's approximation for Bessel functions \cite{Tafeln}, which has been used when deriving Eq. \eqref{eq:E_Gunsenheimer}}. Note that for a given $l$ the spectrum \eqref{eq:E_Gunsenheimer} is the same as in a one-dimensional SNS Josephson junction with the length of the normal region equal to $L(l)$ \cite{deGennes+63,Kulik69JETP}. This is explained by the fact that a quasiparticle with orbital momentum $l$ passes a path of approximately $L(l)$ inside the normal bubble between two consecutive Andreev reflections. Another consequence of this is that the lowest quasiparticle energy in our system is the same as in a SNS junction with the length of the normal region equal to $2a$. We denote this energy, which is by definition the minigap of the system, as $E_g$. It satisfies Eq. \eqref{eq:E_Gunsenheimer} with $n = 0$ and $l = 0$ \cite{Gunsenheimer+96}:
\begin{equation}
	\frac{E_g}{\Delta} \frac{a}{\xi'} = \frac{1}{\pi} \arccos \left( \frac{E_g}{\Delta} \right).
	\label{eq:E_g}
\end{equation}

According to Eq. \eqref{eq:E_Gunsenheimer}, the energy is a monotone function of $l$, since $\partial E_{ln}/ \partial l > 0$ for $E_{ln}>0$. However, when going beyond the Andreev approximation the situation appears to be more complicated. In Ref. \cite{Bespalov+2016} the spectrum of a normal bubble inside a superconductor has been calculated by expanding the quantity $(\Delta - E_{ln})/\Delta$ in powers of $a/\xi'$ in the limit $a \ll \xi'$. In this limit the quantum number $n$ can only take the value $0$ (for $E_{ln}>0$), and the spectrum is
\begin{equation}
	\Delta - E_{l0} \approx 2\Delta \frac{\pi^2}{\xi^{\prime 2} k_F^2} F_l^2(k_F a),
	\label{eq:E_Bespalov}
\end{equation}
\begin{equation}
	 F_l(z) = \int\limits_0^z x^2 j_l^2(x) dx = \frac{z^3}{2} \left[ j_l^2(z) - j_{l+1}(z) j_{l-1}(z) \right],
	\label{eq:F_l}
\end{equation}
where $j_l(z)$ are the spherical Bessel functions. For $l = 0$ and $l = 1$ one has
\begin{equation}
	\Delta-E_{00} = \Delta \frac{\pi^2 a^2}{2\xi^{\prime^2}} \left[ 1 - \frac{\sin(2k_F a)}{2k_F a} \right]^2,
	\label{eq:E0_Bespalov}
\end{equation}
\begin{equation}
	\Delta-E_{10} = \Delta \frac{\pi^2 a^2}{2\xi^{\prime^2}} \left[ 1 + \frac{\sin(2k_F a)}{2k_F a} - 2\frac{\sin^2(k_F a)}{(k_F a)^2} \right]^2.
	\label{eq:E1_Bespalov}
\end{equation}
One can see that the energies of Andreev states are oscillating functions of $a$, which is a consequence of the abrupt order parameter profile, Eq. \eqref{eq:Delta}. Moreover, for large enough $k_F a$ it appears that $E_{0,0} > E_{1,0}$ when $\sin(2k_F a)>0$, which contradicts Eqs. \eqref{eq:E_Gunsenheimer} and \eqref{eq:L(l)}. Thus, Eq. \eqref{eq:E_Gunsenheimer} fails to predict the detailed structure of the subgap spectrum. We suppose that this happens due to the inability of the Andreev approximation to resolve such small energies as $E_{l+1,n} - E_{ln}$, which can be estimated as
\begin{equation}
	\abs{E_{00} - E_{10}} \sim \frac{\Delta^2}{\mu} \frac{a}{\xi'},
	\label{eq:E00-E10}
\end{equation}
according to Eqs. \eqref{eq:E0_Bespalov} and \eqref{eq:E1_Bespalov}.

Another feature that is not captured by the Andreev approximation is the existence of subgap states with $l >k_F a$. In fact, for the order parameter given by Eq. \eqref{eq:Delta} subgap states with arbitrary large $l$ exist \cite{Bespalov+2016}, with their energies approaching the gap edge exponentially fast as $l \to \infty$. It can be seen that the subgap spectrum of a normal bubble is rather complicated even without an impurity. To study the impurity induced states, knowledge of the fine details of the subgap spectrum is not necessary, so we will stick to the quasiclassical approach, which has the form of the Andreev approximation and of the Eilenberger equations \cite{Kopnin-book,EilenbergerEquation} when applied to the BdG equation and to the Green functions, respectively. Within the formalism of the Eilenberger equations the quantum number $l$ changes continuously, and thus the density of states becomes continuous, too. In addition, subgap states with $l >k_F a$ are still not resolved. However, this is not really relevant for the study impurity states (Sec. \ref{sec:Impurity}), since we will make use of Green functions with such arguments (energy and coordinates) that the local density of states vanishes, and hence it is not important whether the spectrum is discrete or continuous. Additionally, we do not consider impurity states with energies close to $\Delta$. Then, the quasiclassical approach is quite reliable.

One more artifact of the Eilenberger equations is the appearance of local minigaps in the density of states, as discussed in Sec. \ref{sub:Clean_Green}.

For now, to determine the global density of states per spin projection, $\nu_g^{(0)}(E)$, we do not need the solution of the Eilenberger equations, as we can make use of Eqs. \eqref{eq:E_Gunsenheimer} and \eqref{eq:L(l)}. Let us start with the general expression:
\begin{equation}
	\nu_g^{(0)}(E) = \sum_{l,n} \sum_{l_z = -l}^l \delta(E - E_{ln}) \int \abs{u_{l,l_z,n}(\vec{r})}^2 d^3 \vec{r}.
	\label{eq:nu_exact}
\end{equation}
Here, $u_{l,l_z,n}(\vec{r})$ is the electron component of the BdG wave function $\psi_{l,l_z,n} = (u_{l,l_z,n}(\vec{r}),v_{l,l_z,n}(\vec{r}))^T$. For the wave functions found in Ref. \cite{Gunsenheimer+96} one has
\begin{equation}
	\int \abs{u_{l,l_z,n}(\vec{r})}^2 d^3 \vec{r} \approx \int \abs{v_{l,l_z,n}(\vec{r})}^2 d^3 \vec{r} \approx \frac{1}{2}.
	\label{eq:int_u2}
\end{equation}
Allowing $l$ to change continuously in Eq. \eqref{eq:nu_exact}, we obtain
\begin{equation}
	\nu_g^{(0)}(E) = \frac{1}{2} \int dl \sum_{n} (2l+1) \delta(E - E_{ln}).
	\label{eq:nu_approx1}
\end{equation}
Here,  $E_{ln}$ can be taken from Eq. \eqref{eq:E_Gunsenheimer}. Note that the quasiclassical approximation is valid only in the limit $k_F a \gg 1$, so that the main contribution to the integral in Eq. \eqref{eq:nu_approx1} comes from $l \gg 1$, hence one can neglect $1/2$ as compared to $l$. For positive energies this yields
\begin{equation}
	\nu_g^{(0)}(E) \approx \int_0^{k_F a} \sum_{n=0}^{\infty} l \delta(E - E_{ln}) dl = \frac{1}{2} \sum_{n=0}^{n_m(E)} \frac{d l_n^2}{dE},
	\label{eq:nu_g1}
\end{equation}
where
\begin{equation}
	n_m(E) = \floor{\frac{E}{\Delta} \frac{a}{\xi'} - \frac{1}{\pi} \gamma(E)},
	\label{eq:n_m}
\end{equation}
\begin{equation}
	l_n^2(E) = k_F^2 a^2 \left\{ 1- \left[ \frac{\Delta \xi'}{\pi a E} \left( \pi n + \gamma(E) \right) \right]^2 \right\},
	\label{eq:l_n(E)}
\end{equation}
and $\floor{x}$ denotes the floor function (the greatest integer less than or equal to $x$). For $n_m(E) =-1$ one has $\nu_g^{(0)}(E) = 0$. In expanded form Eq. \eqref{eq:nu_g1} reads
\begin{eqnarray}
	& \frac{\nu_g^{(0)}(E)}{\nu_0 V} = \frac{3\xi^{\prime 3}}{\pi^2 a^3} \sum\limits_{n=0}^{n_m(E)} \left\{ \frac{\pi n + \arccos \epsilon}{\epsilon^2} \right. & \nonumber\\ 
	&\left. \times \left( \frac{\pi n + \arccos \epsilon}{\epsilon} + \frac{1}{\sqrt{1-\epsilon^2}} \right) \right\},
	\label{eq:nu_g2}
\end{eqnarray}
where $\epsilon = E/\Delta$,  $\nu_0 = mk_F/(2\pi^2 \hbar^2)$ is the normal density of states (per spin projection), and $V = 4\pi a^3/3$ is the volume of the normal region.

\begin{figure}[htb]
	\centering
		\includegraphics[width = \linewidth]{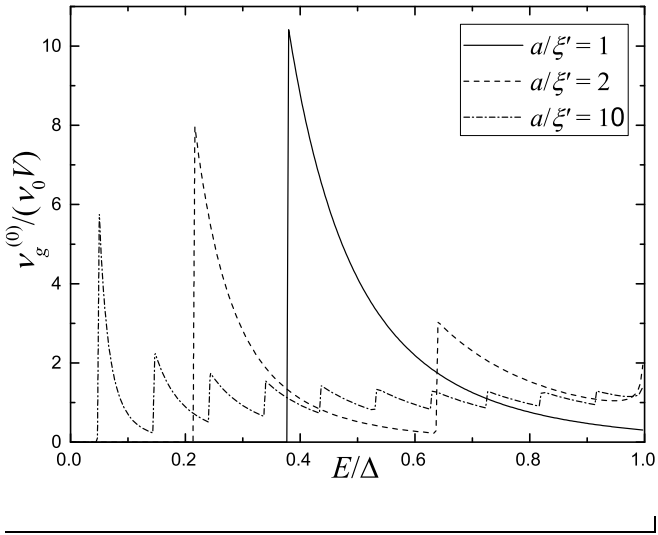}
		\caption{Global density of states in a clean normal bubble [Eq. \eqref{eq:nu_g2}].}
	\label{fig:nuGlobal}
\end{figure}

It can be seen that the density of states [Eq. \eqref{eq:nu_g2}] vanishes for $E<E_g$, and for $E_g < E < \Delta$ it exhibits a sawtooth shape, as can be seen in Fig. \ref{fig:nuGlobal}. The number of peaks on each curve equals $\left\lceil a/\xi' \right\rceil$, where $\left\lceil x \right\rceil$ stands for the ceiling function (the least integer greater than or equal to $x$).

\subsection{Local density of states and Green functions}
\label{sub:Clean_Green}

To determine the local density of states $\nu^{(0)}(E,r)$, the Green function with coinciding coordinates is required. It is known \cite{Bespalov2018} that
\begin{equation}
	G_{ER}^{(0)}(\vec{r},\vec{r}) = \frac{imk_F}{2\pi \hbar^2} \int g_E(\vec{r},\vec{n}) \frac{d^2 \vec{n}}{4\pi},
	\label{eq:GER_qq}
\end{equation}
\begin{equation}
		F_E^{\dagger(0)}(\vec{r},\vec{r}) = \frac{imk_F}{2\pi \hbar^2} \int f^{\dagger}_E(\vec{r},\vec{n}) \frac{d^2 \vec{n}}{4\pi},
	\label{eq:FE_qq}
\end{equation}
where $g_E(\vec{r},\vec{n})$ and $f^{\dagger}_E(\vec{r},\vec{n})$ are quasiclassical Green functions, which satisfy the Eilenberger equations, $\vec{n}$ is a unit vector, and the integrals are over a unit sphere. Explicit expressions for $g_E(\vec{r},\vec{n})$ and $f^{\dagger}_E(\vec{r},\vec{n})$ as well as derivations of further equations in this subsection are given in Appendix \ref{app:Andreev}.

\begin{figure*}[htb]
	\centering
		\includegraphics[width = 0.49\linewidth]{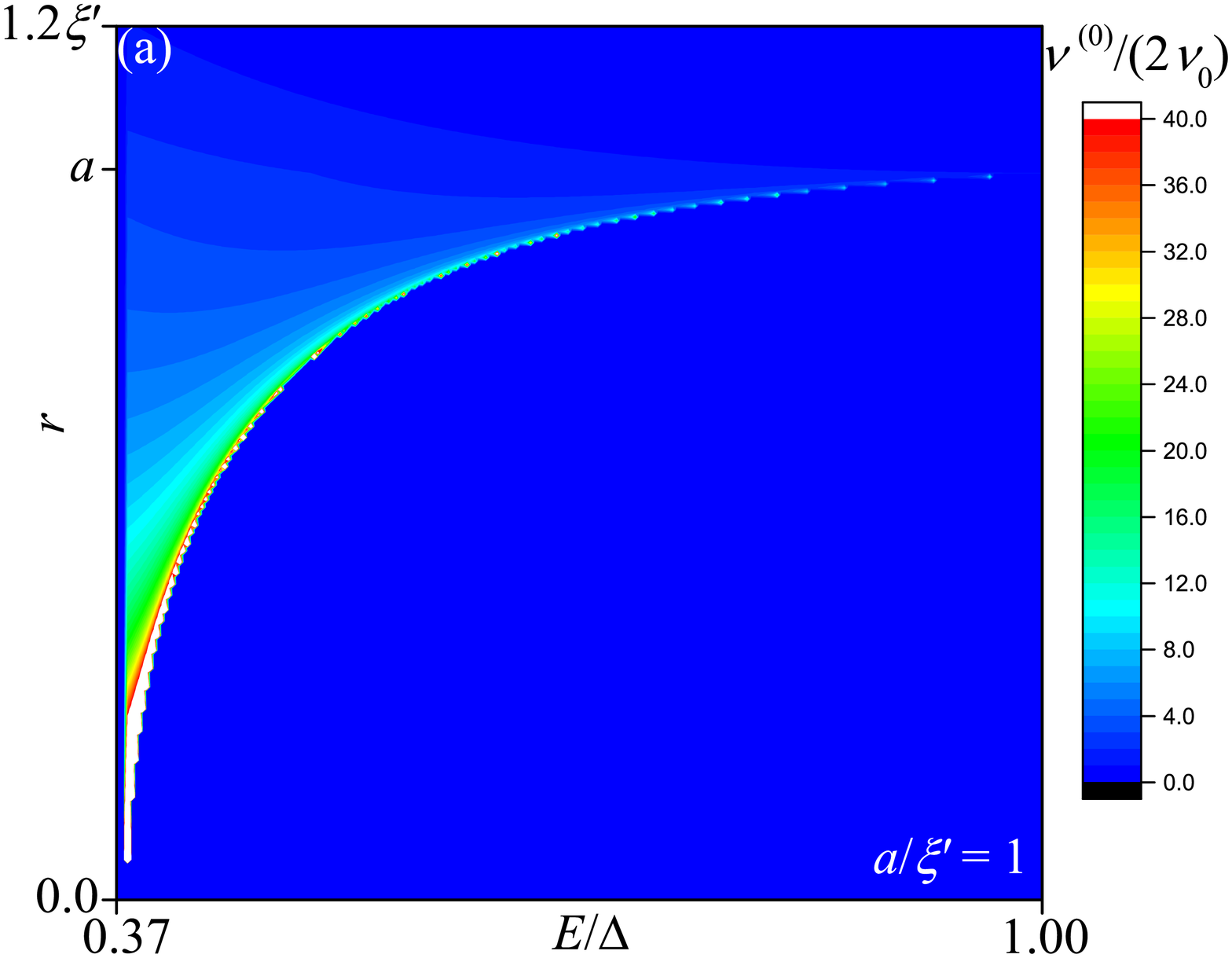}
		\includegraphics[width = 0.49\linewidth]{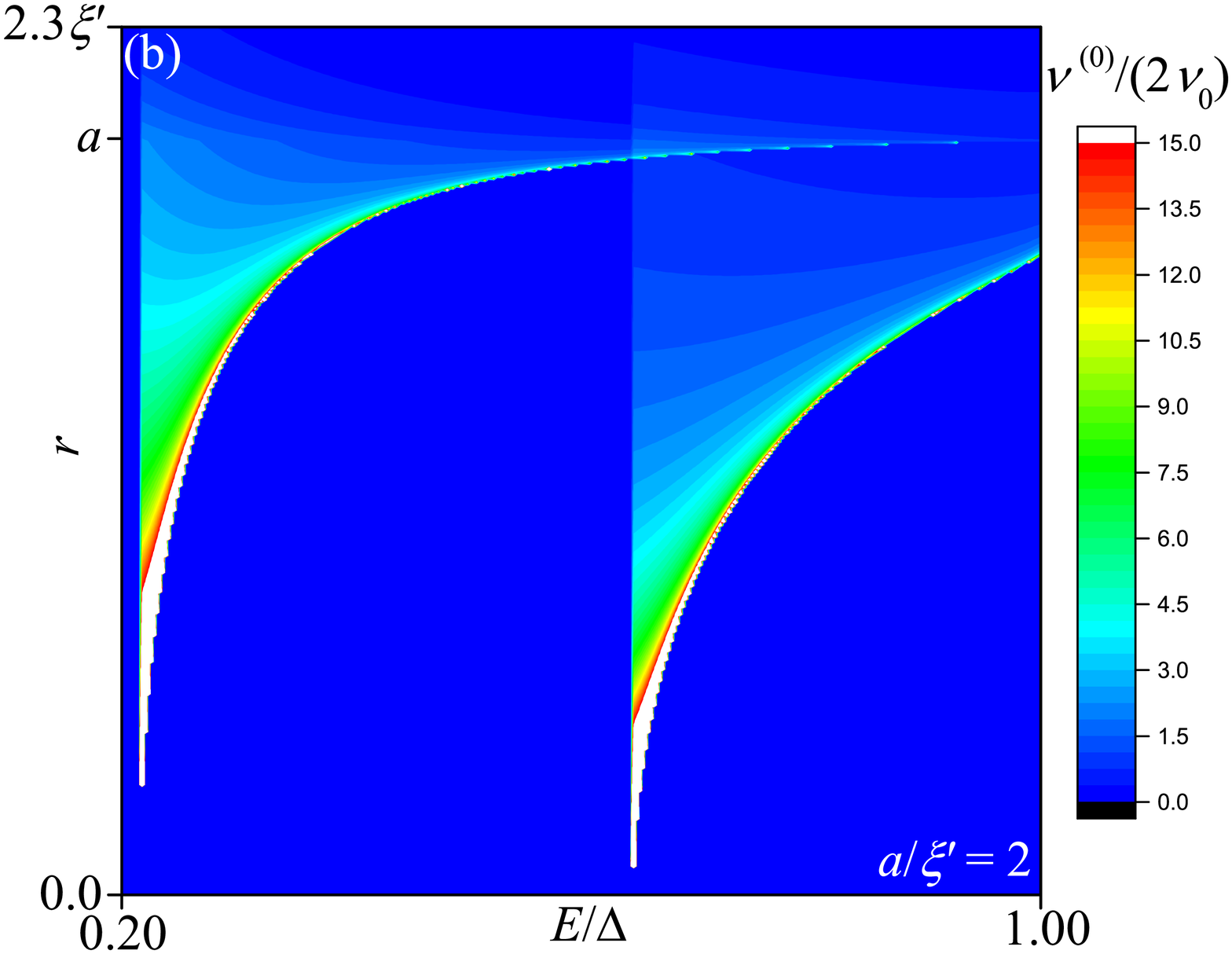}
		\includegraphics[width = 0.49\linewidth]{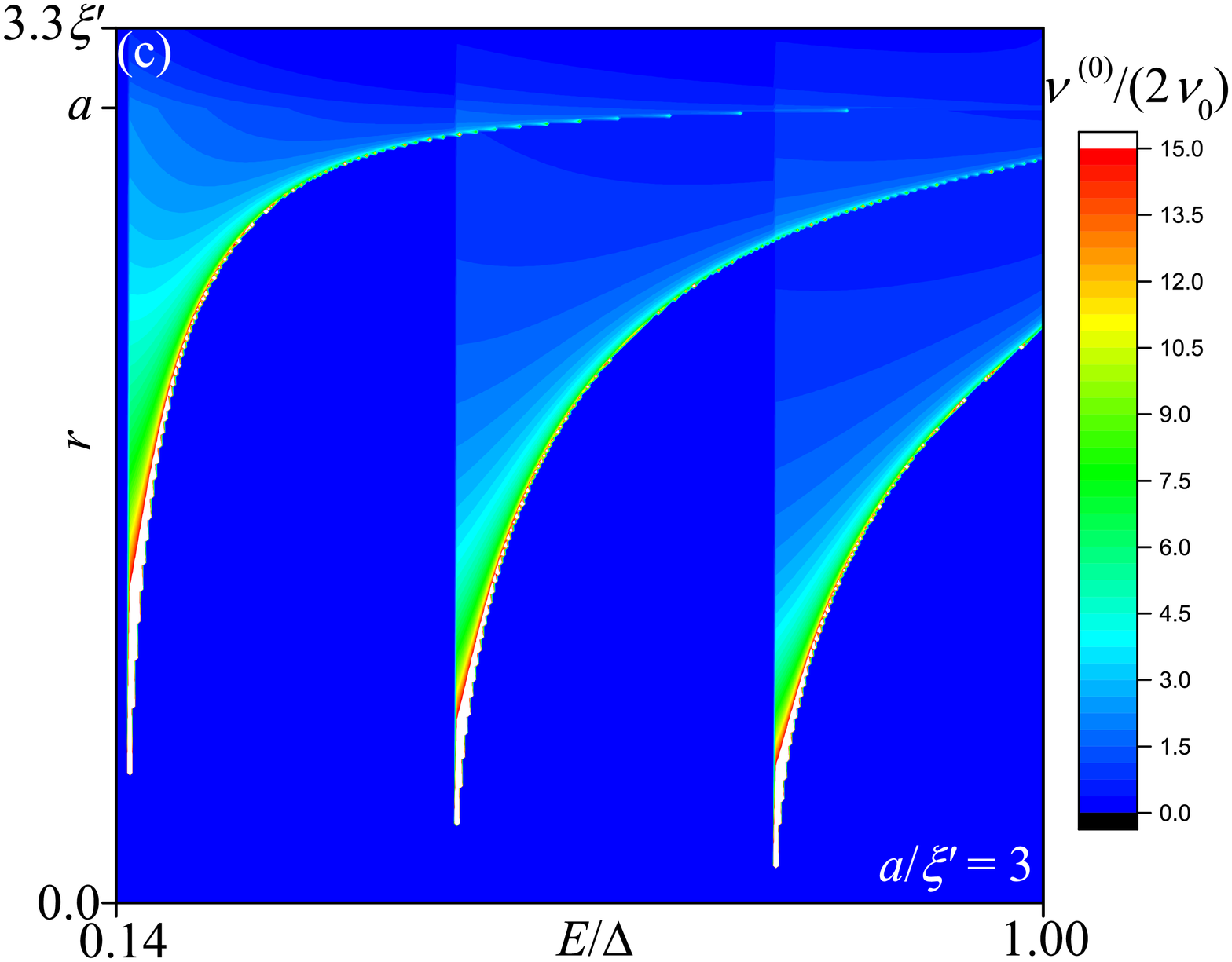}
		\includegraphics[width = 0.49\linewidth]{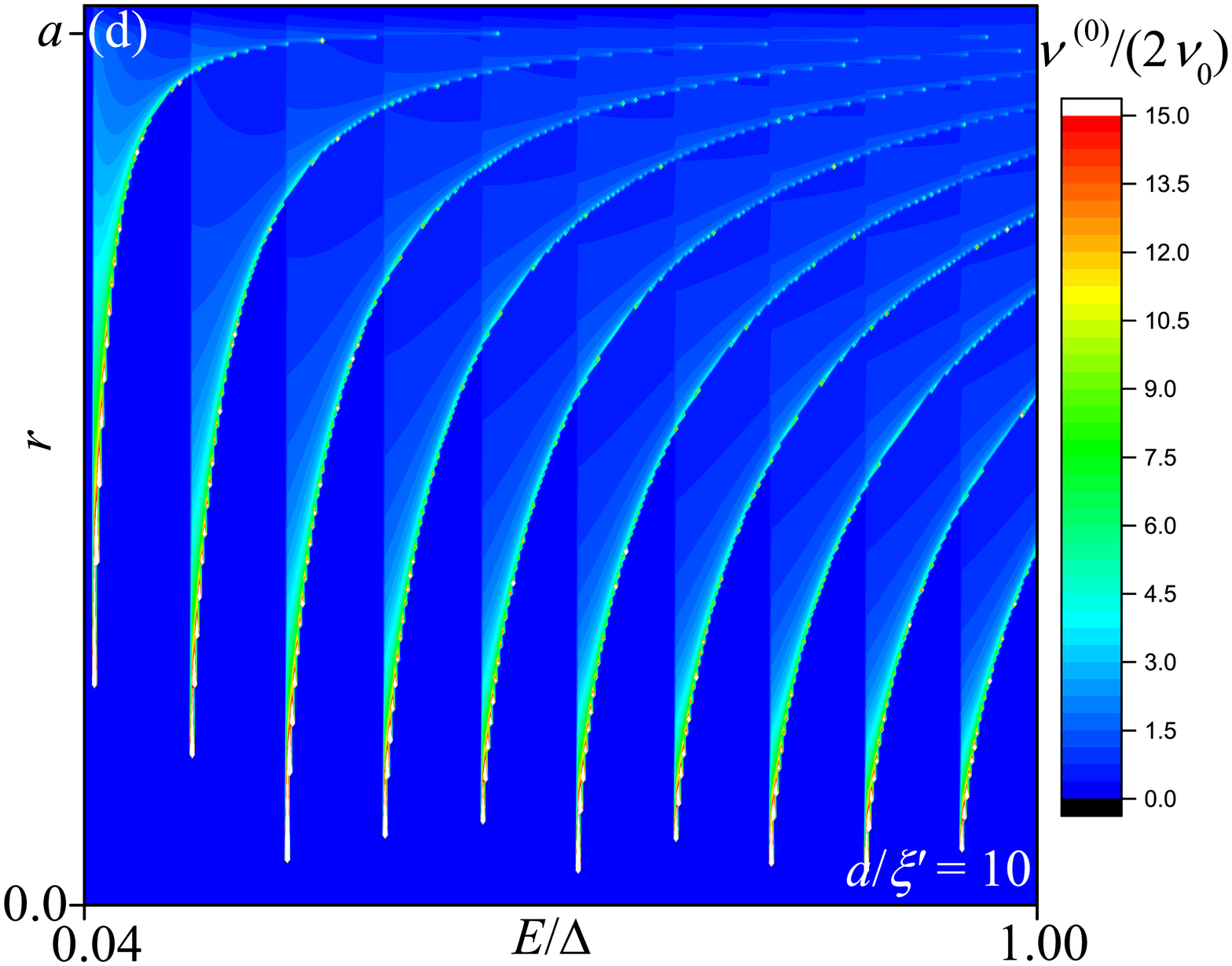}		
	\caption{Local densities of states in the presence of normal bubbles with different radii [Eqs. \eqref{eq:nu_in} and \eqref{eq:nu_out}]: (a) $a = \xi'$, (b) $a = 2\xi'$, (c) $a = 3\xi'$, (d) $a = 10\xi'$.}
	\label{fig:nu}
\end{figure*}

The local density of states is given by the imaginary part of $G_{ER}^{(0)}(\vec{r},\vec{r})$, according to Eq. \eqref{eq:nu_def}. When $E \geq 0$, we find that
\begin{widetext}
\begin{equation}
	\nu^{(0)}(E,r) = \frac{2}{\pi} \mathrm{Im} \left[ G_{ER}^{(0)}(\vec{r},\vec{r}) \right] = \frac{2 \nu_0 \xi'}{\epsilon^2 r} \sum_{n = \floor{\frac{\epsilon}{\xi'} \sqrt{a^2 - r^2} - \frac{\arccos \epsilon}{\pi}} + 1}^{\floor{\epsilon \frac{a}{\xi'} - \frac{\arccos \epsilon}{\pi}}} 
	\frac{n + \frac{\arccos \epsilon}{\pi}}{\sqrt{\frac{r^2 - a^2}{\xi'^2} + \frac{1}{\epsilon^2}\left(n + \frac{\arccos \epsilon}{\pi} \right)^2}}
	\label{eq:nu_in}
\end{equation}
for $r < a$, and
\begin{eqnarray}
	& \nu^{(0)}(E,r) = \frac{2\nu_0 \xi'}{\epsilon^2 r} \sum\limits_{n = 0}^{\floor{\epsilon \frac{a}{\xi'} - \frac{\arccos \epsilon}{\pi}}} \!\!\!\!\!\!
	\frac{n + \frac{\arccos \epsilon}{\pi}}{\sqrt{\frac{r^2 - a^2}{\xi'^2} + \frac{1}{\epsilon^2}\left(n + \frac{\arccos \epsilon}{\pi} \right)^2}} \times & \nonumber \\
	& \times \mathrm{exp}\left\{ -\pi \sqrt{1-\epsilon^2} \left[ \sqrt{\frac{r^2 - a^2}{\xi'^2} + \frac{1}{\epsilon^2}\left(n + \frac{\arccos \epsilon}{\pi} \right)^2} - \frac{1}{\epsilon} \left(n + \frac{\arccos \epsilon}{\pi} \right) \right] \right\} &
	\label{eq:nu_out}
\end{eqnarray}
\end{widetext}
for $r > a$. In Eqs. \eqref{eq:nu_in} and \eqref{eq:nu_out}, if the upper limit of summation is smaller than the lower limit, one should put $\nu^{(0)}(E,r) = 0$. Some profiles on $\nu^{(0)}(E,r)$ are shown in Fig. \ref{fig:nu}.

It can be seen %from Eq. \eqref{eq:nu_in} 
that $\nu^{(0)}(E,r)$ vanishes not only at $E<E_g$, but also at $E>E_g$ for some range of distances $r$:
\begin{equation}
	r < \sqrt{a^2 - \frac{\xi^{\prime 2}}{\epsilon^2} \left( \floor{\epsilon \frac{a}{\xi'} - \frac{\arccos \epsilon}{\pi}} + \frac{\arccos \epsilon}{\pi} \right)^2},
	\label{eq:nu_vanishes}
\end{equation}
which can be derived from Eq. \eqref{eq:nu_in}. The energies sasfying Eq. \eqref{eq:nu_vanishes} lie inside the local minigaps. Let us discuss the origin of this peculiar spectrum. Within the quasiclassical approximation, the density of states in a given point is, roughly speaking, the superposition of local spectra of all classical straight trajectories passing through this point. For a normal inclusion in a superconductor, inside the inclusion on each of these trajectories we have effectively a one-dimensional SNS junction. The subgap spectrum of such junction contains one or several Andreev levels \cite{deGennes+63,Kulik69JETP}. Taking the superposition of these spectra (integrating over $\vec{n}$), we obtain a set of one or several energy bands, which may or may not overlap. For example, in Fig. \ref{fig:nu}c one can see three bands which broaden and begin to overlap as $r$ grows. In the energy range between two neighboring non-overlapping bands the density of states vanishes, which means that we have a local minigap. In addition, there may be a minigap between the band with the highest energy and the bulk gap $\Delta$. This explanation of local minigaps does not rely on spherical symmetry, and hence these spectral features should be common in various systems with locally suppressed superconductivity. However, one can come up with such shapes of normal inclusions in a superconductor that there are no local minigaps except for the one around $E=0$ in any point of space. An example of such shape is a cylinder with a radius of the order of $\xi'$ and a length much larger than $\xi'$.

It should be noted that some features of the obtained density of states [Eqs. \eqref{eq:nu_in} and \eqref{eq:nu_out}] are indeed caused by the spherical symmetry. First, the width of the minigap around $E=0$ does not depend on position in space -- in the general case, this is not so. Second, local minigaps are present for relatively small distances from the center of symmetry. This is explained by the fact that the subgap spectrum in the center is discrete, since the spectra of all classical trajectory passing through the center are the same. If there is at least one Andreev level, there will be a minigap at energies above this level. Specifically for our system, one can see also that the spectral weight in Fig. \ref{fig:nu}a is pushed to the periphery of the bubble when the energy approaches $\Delta$. This effect is relatively easy to explain. 
The spectrum at a distance $r$ from the center of the bubble is the superposition of spectra of one-dimensional SNS junctions with the length of the normal region ranging from $2(a^2-r^2)^{1/2}$ to $2a$. For $a=\xi'$ such junctions host only one (spin-degenerate) Andreev state, whose energy increases and approaches $\Delta$ as the length of the N region decreases. Hence, to obtain a non-zero density of states for energies close $\Delta$ one needs to approach the periphery of the bubble, since only there the spectrum is contributed by trajectories with short enough normal segments.

Strictly speaking, the density of states does not completely vanish inside the local minigaps (though it is exactly equal to zero for $\abs{E}<E_g$). For a given distance $r$ from the origin the density of states inside the minigaps is contributed by Andreev states with orbital momenta $l > k_F r$. The wave functions of these states have exponential tails at $r < k_F^{-1} l$, and thus their contribution to the local density of states and Green functions is exponentially small and can be neglected in our context.

The final ingredients that we will need to analyze the impurity-induced states are the Green functions $G_E^{(0)}(\vec{r},\vec{r}')$ and $F_E^{\dagger (0)}(\vec{r},\vec{r}')$ with noncoinciding coordinates. Here, different expressions exist for $k_F \abs{\vec{r} - \vec{r}'} \lesssim 1 $ and $k_F \abs{\vec{r} - \vec{r}'} \gg 1$ \cite{Bespalov2018}. In our problem all length scales are much larger than the Fermi wavelength, so we are mainly interested in the range of parameters $k_F \abs{\vec{r} - \vec{r}'} \gg 1$, for which the following approximate relations exist \cite{Gorkov+72JETP_eng}:
\begin{widetext}
\begin{equation}
	G_E^{(0)}(\vec{r},\vec{r}') = \frac{m}{2\pi \hbar^2 \abs{\vec{r}-\vec{r}'}} \left[ \tilde{g}_E(\vec{r}',\abs{\vec{r}-\vec{r}'},\vec{n}) e^{ik_F \abs{\vec{r}-\vec{r}'}} - \tilde{g}_E(\vec{r}',-\abs{\vec{r}-\vec{r}'},-\vec{n}) e^{-ik_F \abs{\vec{r}-\vec{r}'}} \right],
	\label{eq:GE(r,ri)}
\end{equation}
\begin{equation}
	F_E^{\dagger(0)}(\vec{r},\vec{r}') = \frac{m}{2\pi \hbar^2 \abs{\vec{r}-\vec{r}'}} \left[ \tilde{f}_E^{\dagger}(\vec{r}',\abs{\vec{r}-\vec{r}'},\vec{n}) e^{ik_F \abs{\vec{r}-\vec{r}'}} - \tilde{f}_E^{\dagger}(\vec{r}',-\abs{\vec{r}-\vec{r}'},-\vec{n}) e^{-ik_F \abs{\vec{r}-\vec{r}'}} \right].
	\label{eq:FE(r,ri)}
\end{equation}
\end{widetext}
Here, $\vec{n} = (\vec{r} - \vec{r}')/\abs{\vec{r}-\vec{r}'}$, and the functions $\tilde{g}_E$ and $\tilde{f}_E^{\dagger}$ satisfy inhomogeneous Andreev equations \eqref{eq:g_Andreev} and \eqref{eq:f_Andreev}.

\section{Impurity-induced states}
\label{sec:Impurity}

\subsection{Impurity states within the minigap}
\label{sub:minigap}

Now we are ready to study the properties of the system with a point impurity. To remind, the impurity is characterized by its position $\vec{r}_i$ and the scattering phases $\alpha_{\uparrow}$ and $\alpha_{\downarrow}$ for spin-up and spin-down electrons, respectively. We have already determined the Green function $G_{E\uparrow\uparrow}(\vec{r},\vec{r}')$ -- see Eqs. \eqref{eq:GE_+impurity} - \eqref{eq:GER_def}. One interesting feature of this function is that it has poles at such energies that ${\cal D}_{\uparrow}(E) = 0$. If these poles appear at real energies, they correspond to localized discrete states. The function ${\cal D}_{\uparrow}(E)$ is generally complex, however it becomes real when $G_{ER}^{(0)}(\vec{r}_i,\vec{r}_i)$ is real. This happens within the global minigap -- at $\abs{E} < E_g$, and also at $E_g< \abs{E} < \Delta$ within local minigaps, where the local density of states vanishes. %$\nu^{(0)}(E,r_i) = 0$. 
The real solutions of ${\cal D}_{\uparrow}(E) = 0$ should be sought in these energy intervals, because otherwise we fall in the continuous part of the energy spectrum, where the appearance of discrete states is very unlikely. In this section we will concentrate on the global minigap, i. e. $\abs{E}<E_g$.

Within the quasiclassical approximation Eq. \eqref{eq:D_up} can be simplified. Indeed, one can see that for a real order parameter $\Delta$ we have
\begin{equation}
	G_{-ER}^{(0)}(\vec{r}_i,\vec{r}_i) = -G_{ER}^{(0)*}(\vec{r}_i,\vec{r}_i),
	\label{eq:G_-ER}
\end{equation}
\begin{equation}
	F_{-E}^{\dagger (0)}(\vec{r}_i,\vec{r}_i) = F_E^{\dagger (0)*}(\vec{r}_i,\vec{r}_i),
	\label{eq:F_-E}
\end{equation}
which follows from Eqs. \eqref{eq:GER_qq}, \eqref{eq:FE_qq}, \eqref{eq:invert_E_g} and \eqref{eq:invert_E_f}. Moreover, for energies lying in the minigap $G_{ER}^{(0)}(\vec{r}_i,\vec{r}_i)$ and $F_E^{\dagger (0)}(\vec{r}_i,\vec{r}_i)$ are real (the imaginary part $i\eps$ of the energy is not relevant then, so that the Gor'kov equation becomes purely real). Hence, for $\abs{E}<E_g$ ${\cal D}_{\uparrow}(E)$ takes the form
\begin{eqnarray}
	& {\cal D}_{\uparrow}(E) = \left[ \frac{mk_F \cot \alpha_{\uparrow}}{2\pi \hbar^2} - G_{ER}^{(0)}(\vec{r}_i,\vec{r}_i) \right] & \nonumber \\
	& \times \left[ \frac{mk_F \cot \alpha_{\downarrow}}{2\pi \hbar^2} + G_{ER}^{(0)}(\vec{r}_i,\vec{r}_i) \right] + F_E^{\dagger (0)2} (\vec{r}_i,\vec{r}_i). &
	\label{eq:D_up_simplified}
\end{eqnarray}
In Appendix \ref{app:ImpStates} we prove that for $\alpha_{\uparrow} \neq 0$, $\alpha_{\downarrow} \neq 0$ and $r_i \neq 0$ the equation ${\cal D}_{\uparrow}(E) = 0$ has two solutions for $\abs{E} < E_g$. Each positive solution corresponds to a spin-up impurity state, and each negative solution corresponds to a spin-down state (if $G_{E\uparrow\uparrow}(\vec{r},\vec{r}')$ has a pole at $E = E_0$, then $G_{E\downarrow\downarrow}(\vec{r},\vec{r}')$ has a pole at $E = -E_0$). Thus, an impurity generally induces two discrete subgap states, even if it is nonmagnetic. The appearance of impurity states inside the minigap is closely related to a similar phenomenon in narrow SNS junctions. Indeed, such junctions are one-dimensional analogues of our system, if there is no phase difference between the superconducting banks. The subgap spectrum of such junction consists of discrete Andreev states. The effect of a nonmagnetic point impurity (barrier) on these states has been studied in Refs. \cite{Zaikin+80JETP,Bagwell92PRB}. It has been found that a point impurity generally lowers the energy of the the lowest Andreev state, unless the defect is placed exactly in the center of the junction -- in this case, the subgap spectrum remains unchanged. It can be seen that our three-dimensional system exhibits very similar behavior. In this context it is also worth mentioning an analogous result obtained by Liu at al. \cite{Liu+2018PRB}, who predicted impurity-induced subgap states in a superconductor-normal metal heterostructure.

Returning to our system, for a given pair of scattering phases $\alpha_{\uparrow}$ and $\alpha_{\downarrow}$ we can figure out the number of positive and negative solutions of ${\cal D}_{\uparrow}(E) = 0$. According to considerations from Appendix \ref{app:ImpStates}, ${\cal D}_{\uparrow}(E) = 0$ has solutions with both signs when ${\cal D}_{\uparrow}(0) > 0$ [then ${\cal D}_{\uparrow+}(0)$ and ${\cal D}_{\uparrow-}(0)$, given by Eq. \eqref{eq:D+-}, have opposite signs]. The quantity ${\cal D}_{\uparrow}(0)$ is easy to evaluate:
\begin{equation}
	{\cal D}_{\uparrow}(0) = \left( \frac{mk_F}{2\pi\hbar^2} \right)^2 \left( 1 + \cot \alpha_{\uparrow} \cot \alpha_{\downarrow} \right).
	\label{eq:D(0)}
\end{equation}
Hence, the condition for the existence of impurity states with opposite spins has the form
\begin{equation}
	\sin \alpha_{\uparrow} \sin \alpha_{\downarrow} > 0 \quad \mbox{or} \quad \cos(\alpha_{\uparrow} - \alpha_{\downarrow}) < 0.
	\label{eq:spin_up+down}
\end{equation}

Consider the range of parameters
\begin{equation}
	\alpha_{\uparrow} > 0, \quad \alpha_{\downarrow} < 0, \quad \cos(\alpha_{\uparrow} - \alpha_{\downarrow}) > 0.
	\label{eq:2spin_up}
\end{equation}
Here, we have two spin-up impurity states [because ${\cal D}_{\uparrow \pm}(0) <0$]. Correspondingly, for the parameters
\begin{equation}
	\alpha_{\uparrow} < 0, \quad \alpha_{\downarrow} > 0, \quad \cos(\alpha_{\uparrow} - \alpha_{\downarrow}) > 0
	\label{eq:2spin_down}
\end{equation}
there are two spin-down states.

\begin{figure}[htb]
	\centering
		\includegraphics[width=\linewidth]{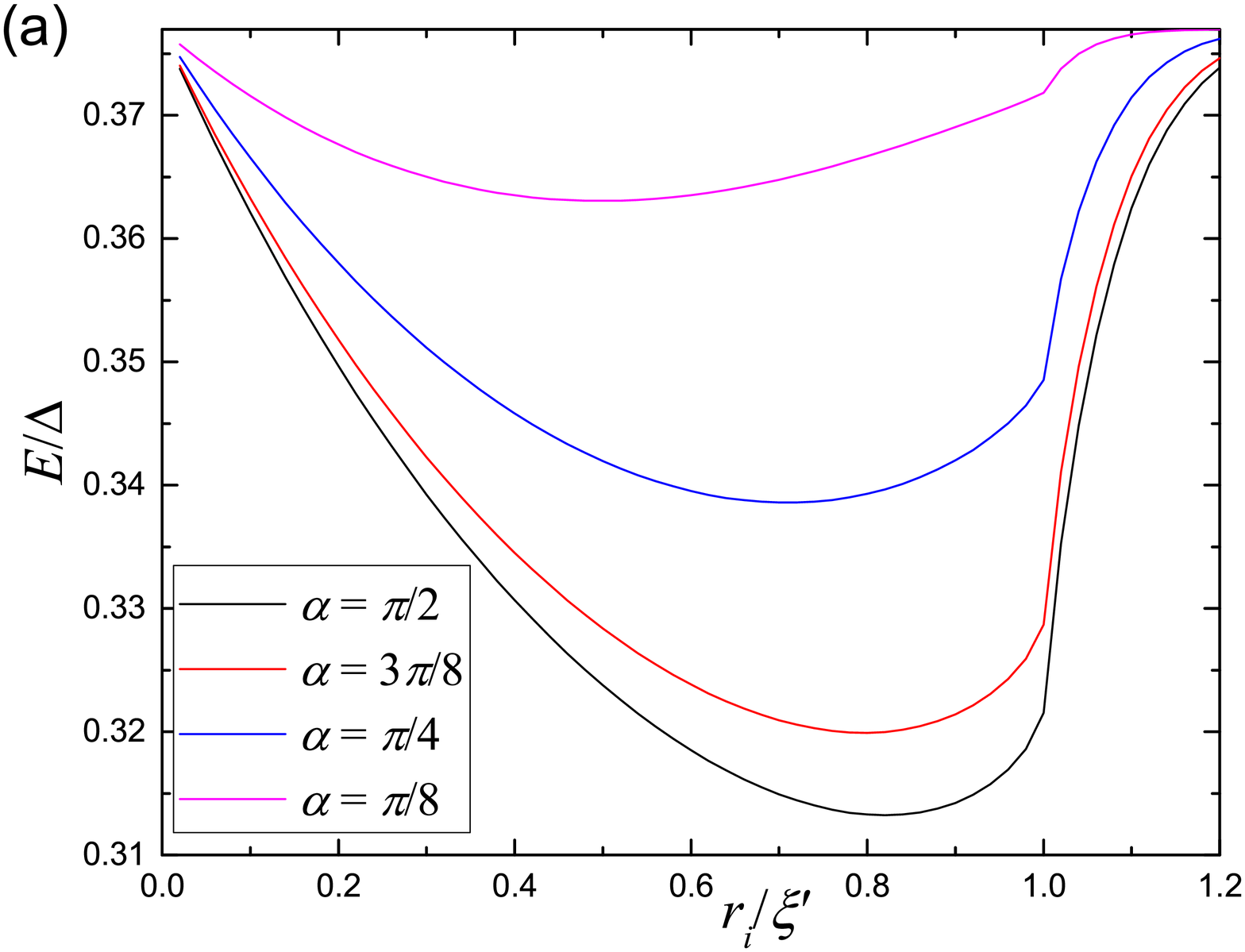}
		\includegraphics[width=\linewidth]{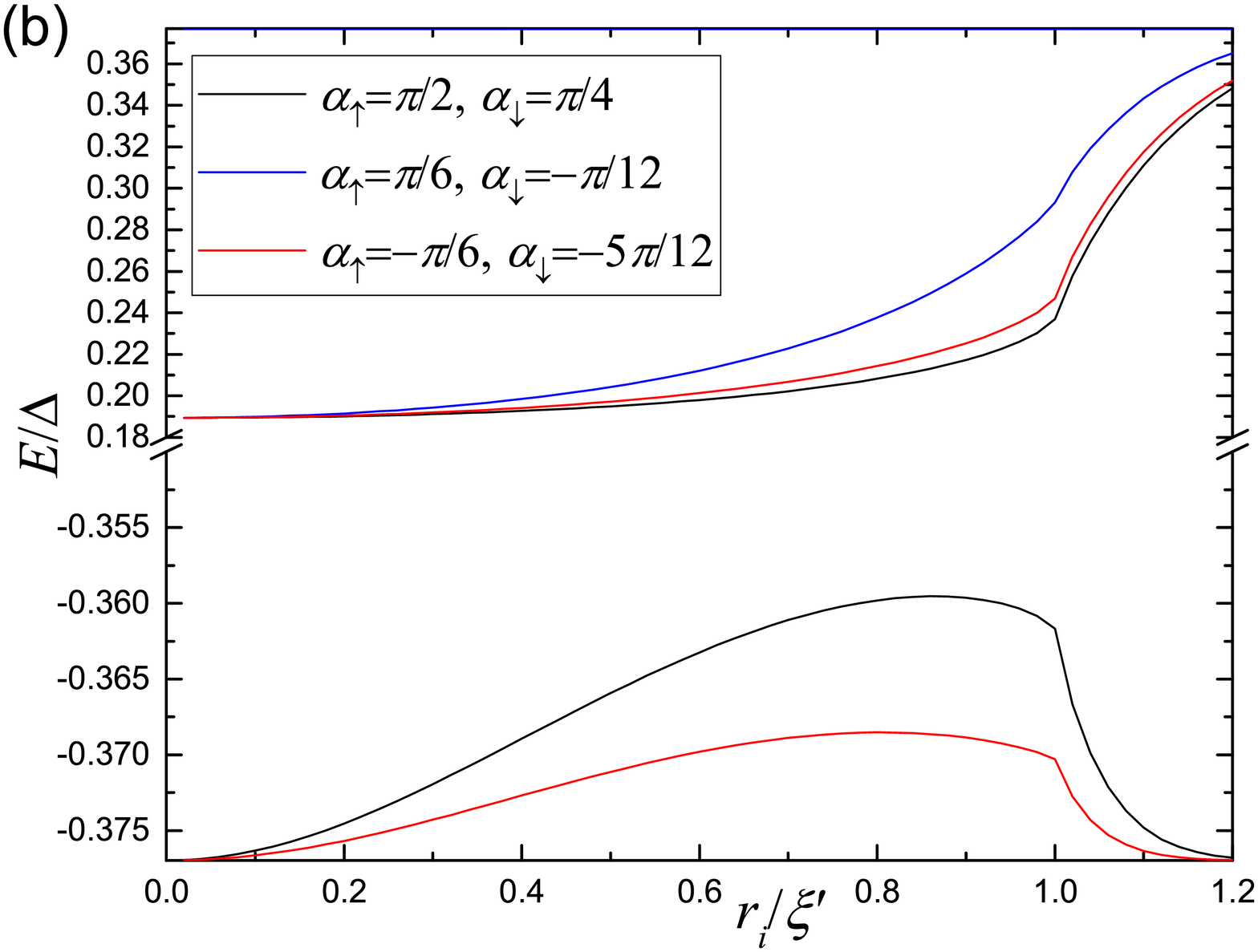}
	\caption{Energies of spin-up impurity states [solutions of ${\cal D}_{\uparrow}(E) = 0$] vs. $r_i$. We take $a = \xi'$, so that $E_g = 0.377 \Delta$. The upper boundary of the graphs corresponds to $E = E_g$. (a) The impurity is nonmagnetic: $\alpha_{\uparrow} = \alpha_{\downarrow} = \alpha$. For each positive solution of ${\cal D}_{\uparrow}(E) = 0$ there is also a negative solution with the same modulus, which is not shown. (b) The case of a magnetic impurity. The upper curve for $\alpha_{\uparrow} = \pi/6$, $\alpha_{\downarrow} = -\pi/12$ lies very close to $E_g$.}
	\label{fig:Evsr}
\end{figure}

Some dependencies of the solutions of ${\cal D}_{\uparrow}(E) = 0$ vs. $r_i$ are shown in Fig. \ref{fig:Evsr}.

In some special cases, there may be only one impurity state. For example, let us take $\alpha_{\uparrow} \neq 0$, $\alpha_{\downarrow} = 0$. Then, to find the spin-up impurity states one should solve
\begin{equation}
	\frac{mk_F \cot \alpha_{\uparrow}}{2\pi \hbar^2} - G_{ER}^{(0)}(\vec{r}_i,\vec{r}_i) = 0.
	\label{eq:alpha_down=0}
\end{equation}
This equation has one solution, because of the monotony of the $G_{ER}^{(0)}(\vec{r}_i,\vec{r}_i)$ vs. $E$ dependence [see Eq. \eqref{eq:F'<G'}]. This solution is positive for $\alpha_{\uparrow} > 0$, and hence there is a spin-up impurity state. For $\alpha_{\uparrow} < 0$ there is a spin-down state. If one puts $\alpha_{\uparrow} = \alpha_{\downarrow} = 0$, one has effectively no impurity and hence no impurity states.

Another special case is when $r_i = 0$, i. e. the impurity is in the center of the normal bubble. Then 
\begin{eqnarray}
	& {\cal D}_{\uparrow}(E) = \left( \frac{mk_F}{2\pi \hbar^2} \right) \left[ 1 + \cot \left(\gamma - \frac{2Ea}{\hbar v_F} \right) \left( \cot \alpha_{\uparrow} - \cot \alpha_{\downarrow} \right) \right. & \nonumber \\
	& \left. + \cot \alpha_{\uparrow} \cot \alpha_{\downarrow} \right]. &
	\label{eq:D_ri=0}
\end{eqnarray}
The equation ${\cal D}_{\uparrow}(E) = 0$ can be reduced to
\begin{equation}
	\cot\left( \gamma(E) - \frac{2Ea}{\hbar v_F} \right) = \cot (\alpha_{\uparrow} - \alpha_{\downarrow}).
	\label{eq:E_up_ri=0}
\end{equation}
For $\cot (\alpha_{\uparrow} - \alpha_{\downarrow}) \neq 0$ this equation has one solution, because the left-hand side is monotonous in $E$ and it takes all real values when $E \in (-E_g,E_g)$. For $\cot (\alpha_{\uparrow} - \alpha_{\downarrow}) > 0$ we have a spin-up impurity state, and for $\cot (\alpha_{\uparrow} - \alpha_{\downarrow}) < 0$ there is a spin-down state. When $\tan (\alpha_{\uparrow} - \alpha_{\downarrow}) = 0$ (nonmagnetic impurity), there are no impurity states.

Now we will analyze the local structure of the density of states at $E \in (-E_g,E_g)$. Let us consider the most common case when there are two impurity states. We denote as $E_{1\uparrow}$ and $E_{2\uparrow}$ the solutions of ${\cal D}_{\uparrow}(E) = 0$, and as $(u_{1\uparrow}(\vec{r}),v_{1\uparrow}(\vec{r}))^T$ and $(u_{2\uparrow}(\vec{r}),v_{2\uparrow}(\vec{r}))^T$ the corresponding normalized solutions of the BdG equations. In the general case the quasiparticle wave functions have also spin-down components, $u_{\downarrow}(\vec{r})$ and $v_{\downarrow}(\vec{r})$, however, in our situation they vanish. The density of states for spin-up electrons at $\abs{E}<E_g$ is
\begin{equation}
	\nu_{\uparrow}(E,\vec{r}) = \delta(E - E_{1 \uparrow}) \abs{u_{1\uparrow}(\vec{r})}^2 + \delta(E - E_{2 \uparrow}) \abs{u_{2\uparrow}(\vec{r})}^2.
	\label{eq:nu_up}
\end{equation}
The function $G_{E\downarrow\downarrow}(\vec{r},\vec{r}')$ has poles at $E_{1\downarrow} = -E_{1\uparrow}$ and $E_{2\downarrow} = -E_{2\uparrow}$. We denote the corresponding wave functions of spin-down quasiparticles as $(u_{1\downarrow},v_{1\downarrow})^T$ and $(u_{2\downarrow},v_{2\downarrow})^T$. These functions can be chosen in such a way that $u_{1\downarrow} = v_{1\uparrow}$ and $u_{2\downarrow} = v_{2\uparrow}$ (also note that all wave functions can be chosen real). Then the density of states of spin-down electrons is
\begin{equation}
	\nu_{\downarrow}(E,\vec{r}) = \delta(E + E_{1\uparrow}) \abs{v_{1\uparrow}(\vec{r})}^2 + \delta(E + E_{2\uparrow}) \abs{v_{2\uparrow}(\vec{r})}^2.
	\label{eq:nu_down}
\end{equation}

\begin{figure*}[htb]
	\centering
		\includegraphics[width=0.49\linewidth]{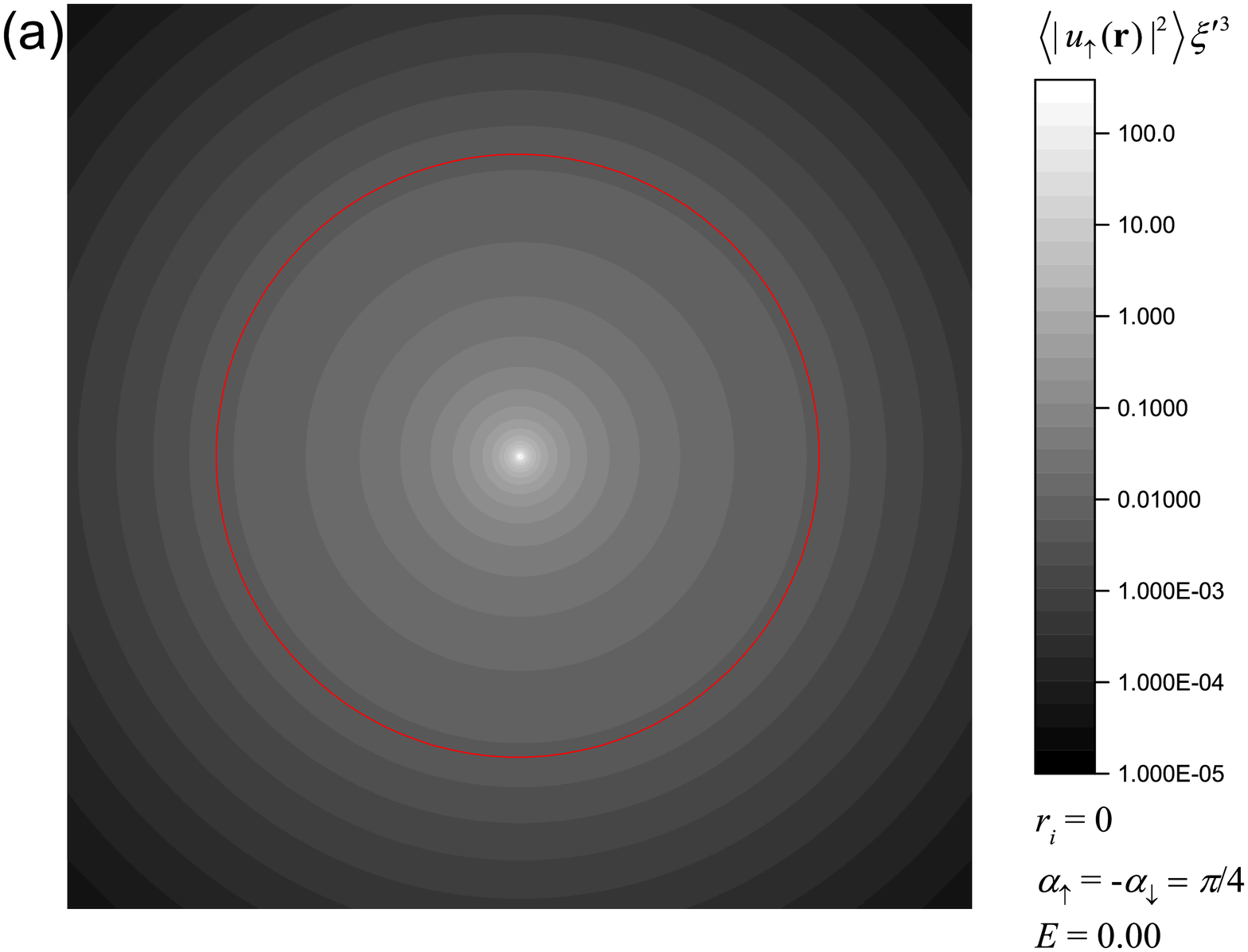}
		\includegraphics[width=0.49\linewidth]{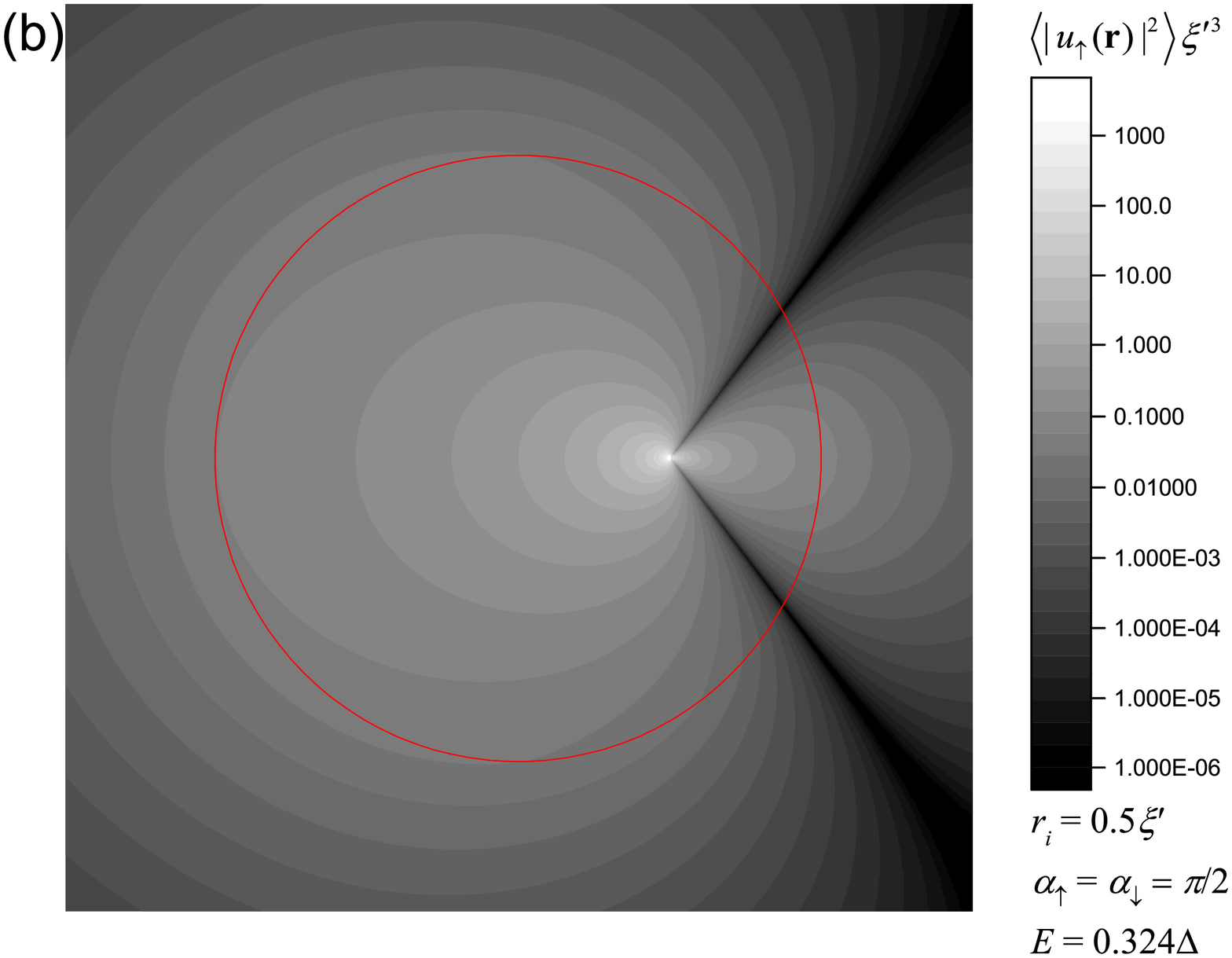} 
		\includegraphics[width=0.49\linewidth]{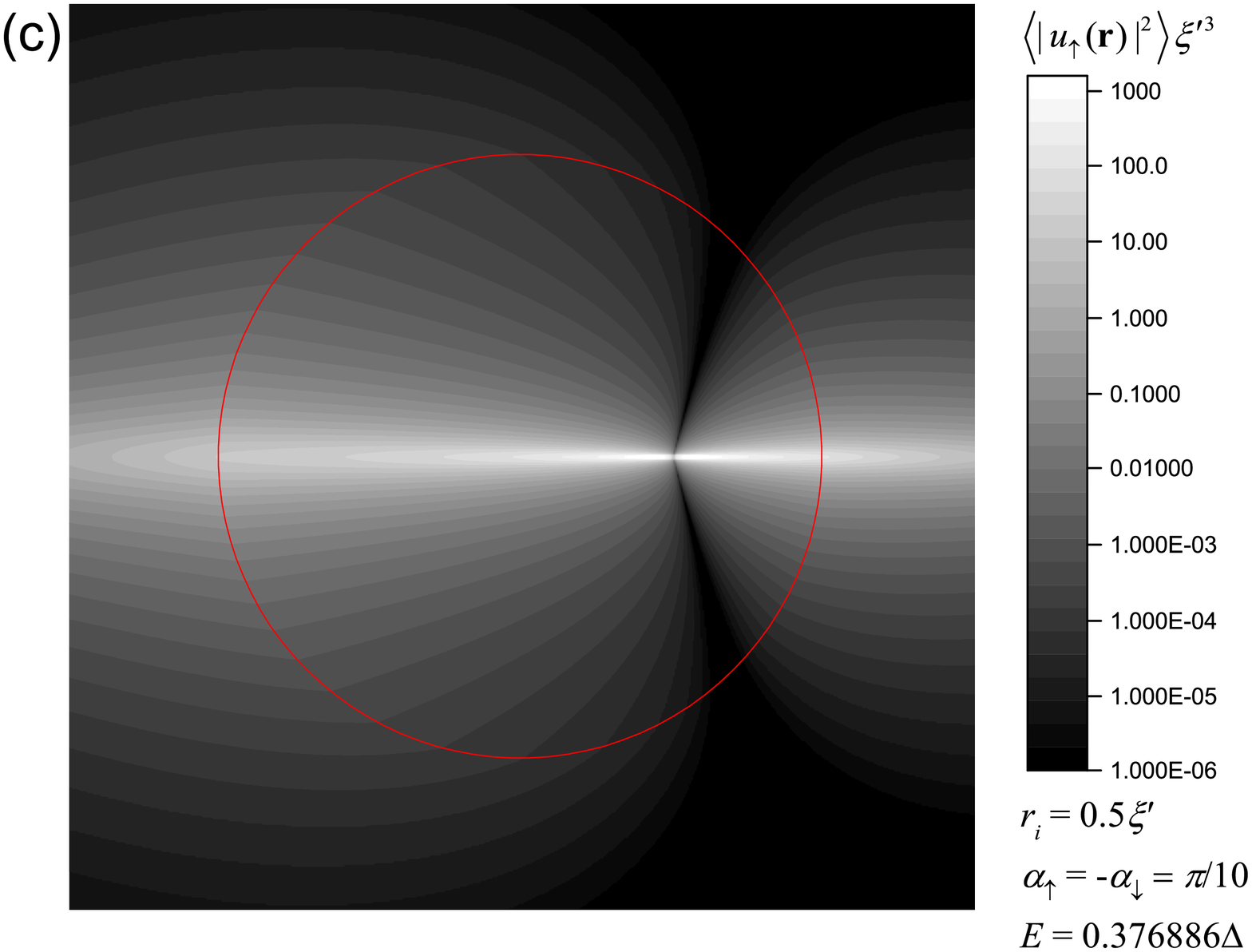}
		\includegraphics[width=0.49\linewidth]{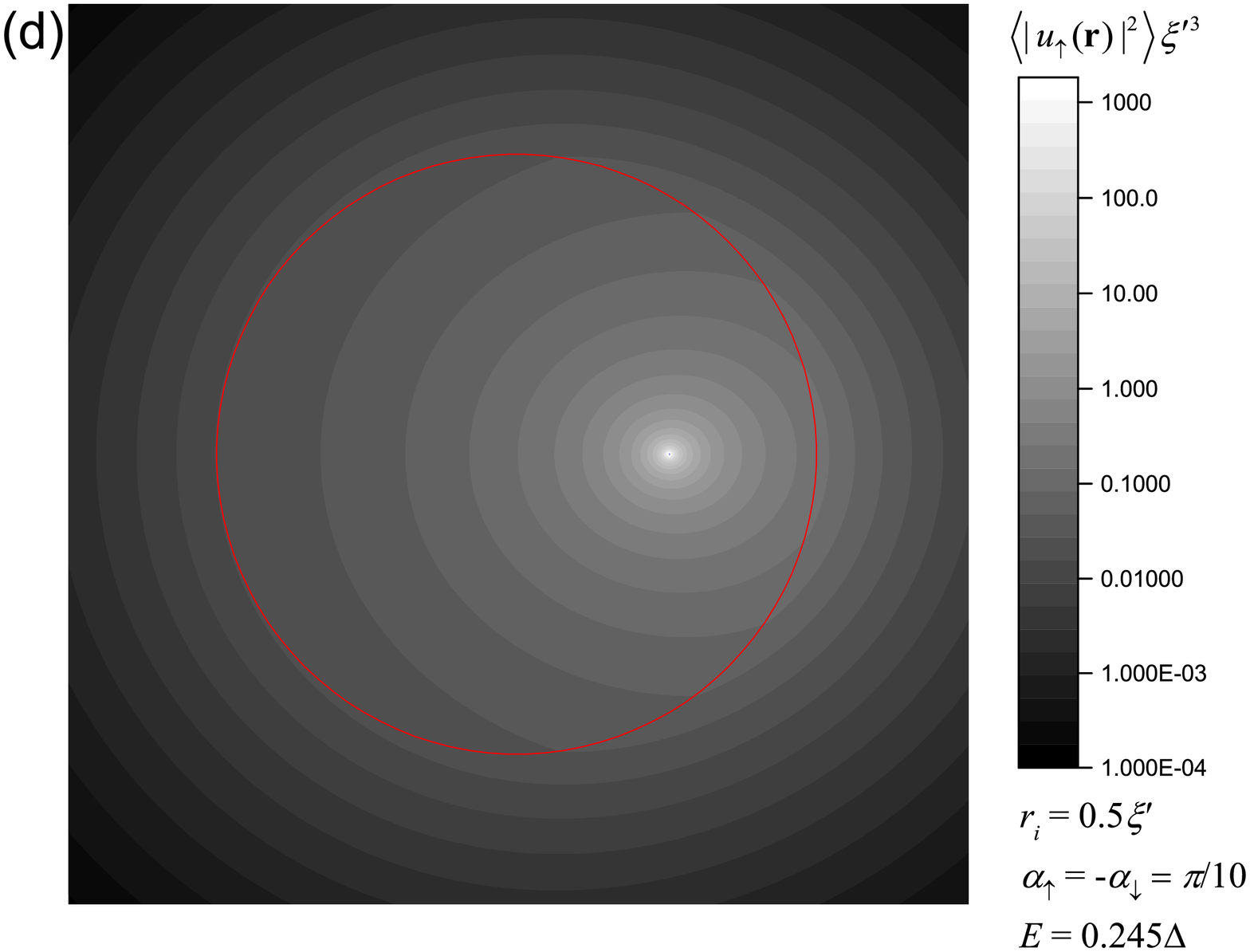}
		\includegraphics[width=0.49\linewidth]{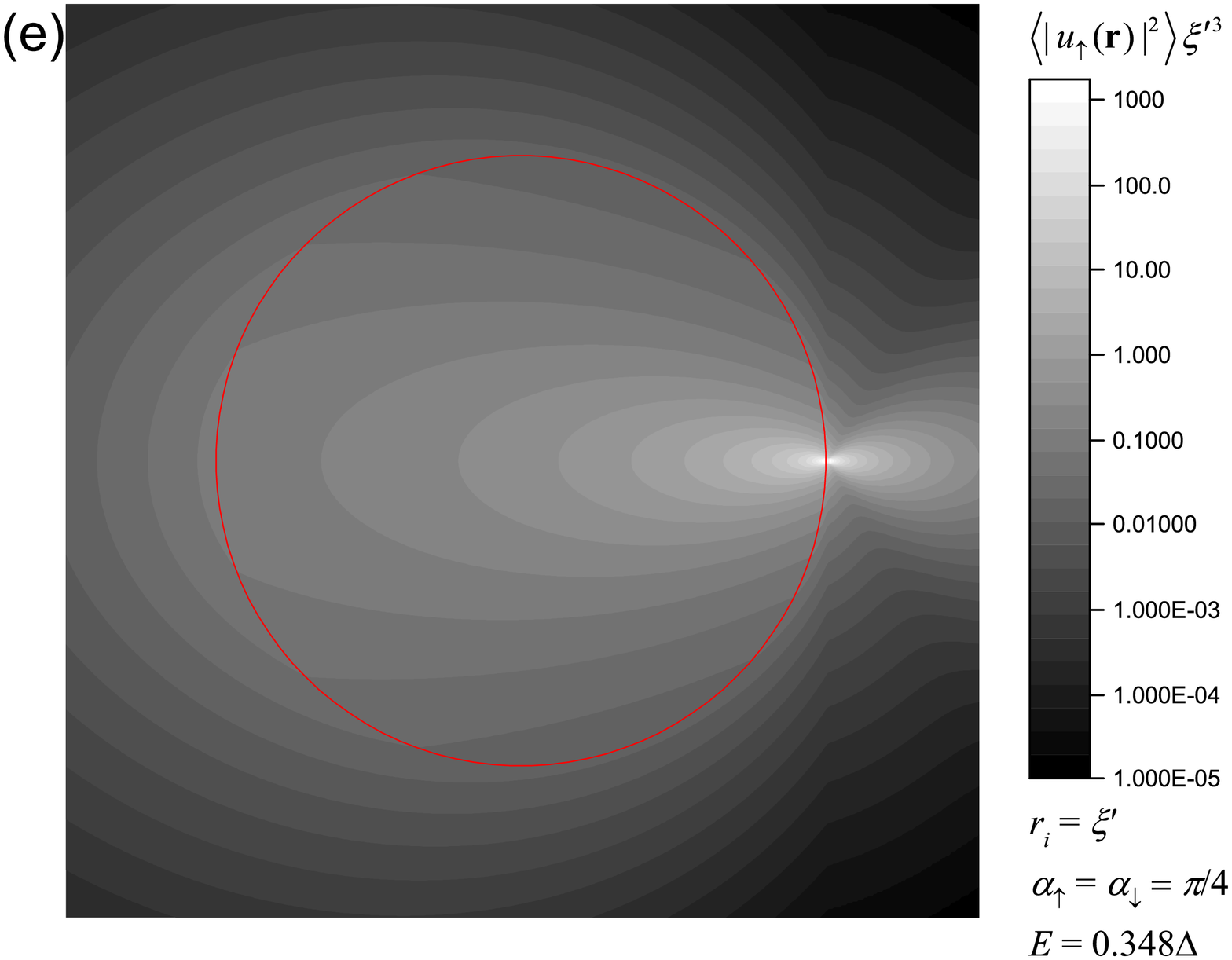}
		\includegraphics[width=0.49\linewidth]{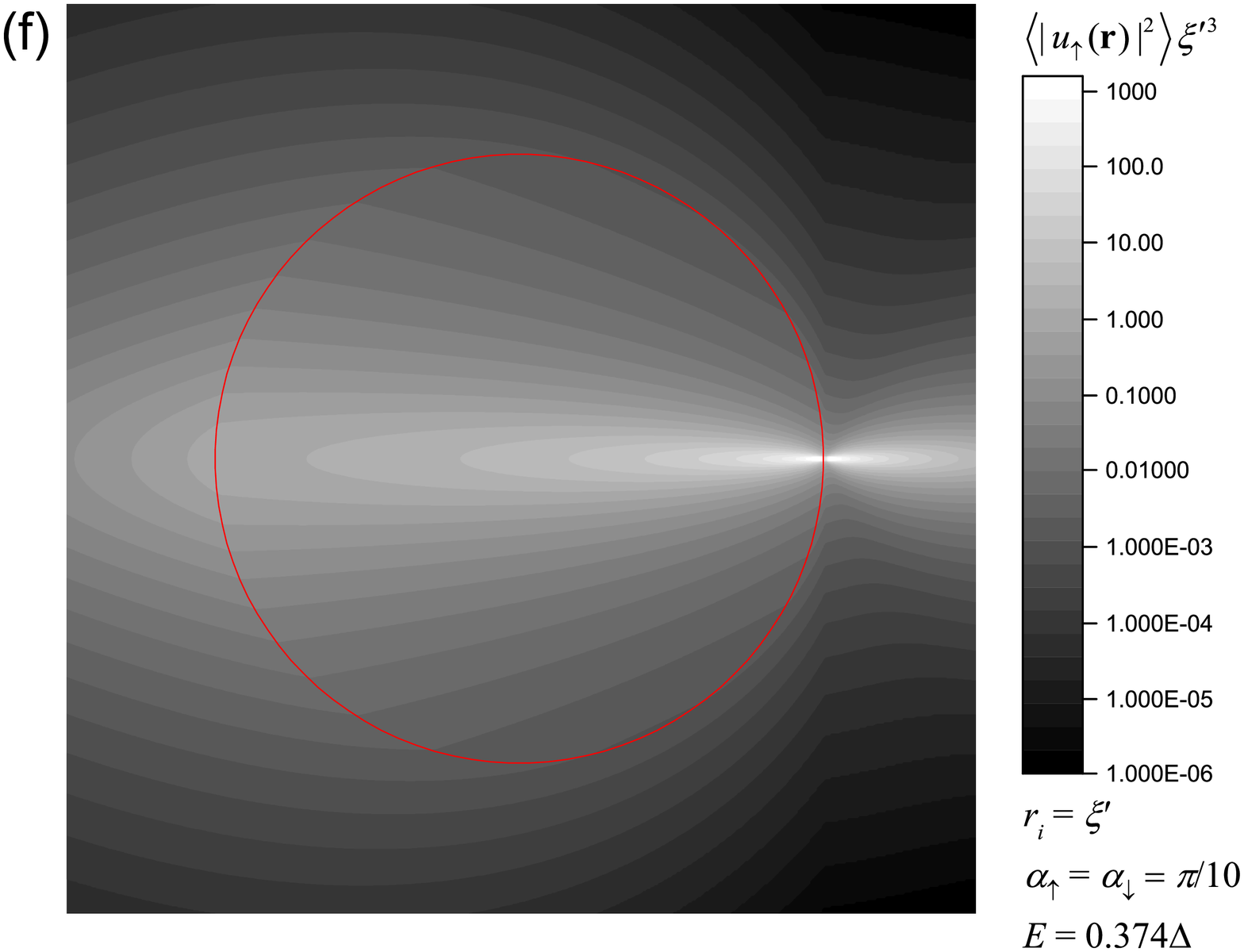}
	\caption{Profiles of impurity states in a plane passing through the center of the normal bubble and through the impurity. $u_{\uparrow}$ stands for either $u_{1\uparrow}$ or $u_{2\uparrow}$ (or both, if the impurity is nonmagnetic). The normal bubble is circled in red and has a ridius $a = \xi'$. Dimensions of the area shown are $3\xi' \times 3\xi'$. Characteristics of the impurity -- $r_i$, $\alpha_{\uparrow}$ and $\alpha_{\downarrow}$ -- and the energies $E$ of impurity states are written in each graph.}
	\label{fig:ImpState}
\end{figure*}

The explicit form of $u_{1\uparrow}(\vec{r})$ and $v_{1\uparrow}(\vec{r})$ is given by Eqs. \eqref{eq:u_imp} - \eqref{eq:v_imp}. For $k_F \abs{\vec{r} - \vec{r}_i} \gg 1$, using Eqs. \eqref{eq:GE(r,ri)}, \eqref{eq:FE(r,ri)} and \eqref{eq:invert_n_g} - \eqref{eq:invert_E_f} we can rewrite $u_{1\uparrow}(\vec{r})$ and $v_{1\uparrow}(\vec{r})$ as follows:
\begin{widetext}
\begin{equation}
	u_{1\uparrow}(\vec{r}) = \frac{m e^{ik_F \abs{\vec{r} - \vec{r}_i}}}{2 \pi \hbar^2 \abs{\vec{r} - \vec{r}_i}} \left[A_{\uparrow} \tilde{g}_E(\vec{r}_i,\abs{\vec{r} - \vec{r}_i},\vec{n}) - B_{\uparrow} \tilde{f}_E^{\dagger}(\vec{r}_i,\abs{\vec{r} - \vec{r}_i},\vec{n}) \right] + \mathrm{c.c.},
	\label{eq:u_up}
\end{equation}
\begin{equation}
	v_{1\uparrow}(\vec{r}) = \frac{m e^{ik_F \abs{\vec{r} - \vec{r}_i}}}{2 \pi \hbar^2 \abs{\vec{r} - \vec{r}_i}} \left[A_{\uparrow} \tilde{f}_E^{\dagger}(\vec{r}_i,\abs{\vec{r} - \vec{r}_i},\vec{n}) + B_{\uparrow} \tilde{g}_{-E}^{\dagger}(\vec{r}_i,\abs{\vec{r} - \vec{r}_i},\vec{n}) \right] + \mathrm{c.c.},
	\label{eq:v_up}
\end{equation}
where $\vec{n} = (\vec{r}- \vec{r}_i)/\abs{\vec{r}- \vec{r}_i}$, and c.c. stands for the complex conjugate. It can be seen that the wave functions oscillate on the scale $2\pi k_F^{-1}$, and thus the density of states has ripples. Below we will plot the density of states averaged over an oscillation period, bearing in mind that this average also gives the amplitude of the ripples. We denote the spatially averaged quantities as $\mean{...}$. For the wave functions we have
\begin{equation}
	\mean{\abs{u_{1\uparrow}(\vec{r})}^2} = 2 \left( \frac{m}{2\pi \hbar^2 \abs{\vec{r} - \vec{r}_i}} \right)^2 \abs{A_{\uparrow} \tilde{g}_E(\vec{r}_i,\abs{\vec{r}-\vec{r}_i},\vec{n}) - B_{\uparrow} \tilde{f}^{\dagger}_{-E}(\vec{r}_i,\abs{\vec{r}-\vec{r}_i},\vec{n})}^2,
	\label{eq:mean_u2}
\end{equation}
\begin{equation}
	\mean{\abs{v_{1\uparrow}(\vec{r})}^2} = 2 \left( \frac{m}{2\pi \hbar^2 \abs{\vec{r} - \vec{r}_i}} \right)^2 \abs{A_{\uparrow} \tilde{f}^{\dagger}_E(\vec{r}_i,\abs{\vec{r}-\vec{r}_i},\vec{n}) + B_{\uparrow} \tilde{g}_{-E}(\vec{r}_i,\abs{\vec{r}-\vec{r}_i},\vec{n})}^2.
	\label{eq:mean_v2}
\end{equation}
\end{widetext}
It follows from Eqs. \eqref{eq:invert_E_g}, \eqref{eq:invert_E_f} and \eqref{eq:abs_g=abs_f} that $\mean{\abs{u_{1\uparrow}(\vec{r})}^2} = \mean{\abs{v_{1\uparrow}(\vec{r})}^2}$, and one can prove the same relation for $u_{2\uparrow}(\vec{r})$ and $v_{2\uparrow}(\vec{r})$. This proves that
\begin{equation}
	\mean{\nu_{\downarrow}(E,\vec{r})} = \mean{\nu_{\uparrow}(-E,\vec{r})},
	\label{eq:nu_up_down}
\end{equation}
and hence it is sufficient to plot $\mean{\abs{u_{1\uparrow}(\vec{r})}^2}$ and $\mean{\abs{u_{2\uparrow}(\vec{r})}^2}$ to get an understanding of the behavior of the density of states. Some profiles of these functions are shown in Fig. \ref{fig:ImpState}.

To conclude, we note that the existence of two impurity states, as demonstrated in Appendix \ref{app:ImpStates}, follows from some quite general analytical properties of the Green functions with coinciding arguments and of ${\cal D}_{\uparrow}(E)$. It turns out that these properties hold for an arbitrary superconductor with a real spatially inhomogeneous gap in the absence of a magnetic field, if the impurity is placed in a position that is not a center of inversion symmetry for $\Delta(\vec{r})$. This statement in proved in Appendix \ref{app:realistic}. Thus, the emergence of two localized states induced by a point impurity should be quite common, and our main qualitative results hold for a normal sphere inside a superconductor with a realistic (self-consistent) order parameter profile.

\subsection{Impurity states outside the minigap}
\label{sub:Above_Eg}

Remarkably, discrete impurity states appear not only inside the minigap, but also in the continuous spectrum, i.e. at $E > E_g$. This is possible because of the existence of local minigaps, where the local density of states vanishes and the Green functions become real. It is proved in Appendix \ref{app:ImpStates} that for a given impurity position $\vec{r}_i$ the defect induces from two to four discrete states inside each local minigap (i.e. in each energy interval where $\nu^{(0)}(E,r_i) = 0$). A numerical solution of the equation ${\cal D}_{\uparrow}(E)$ with different parameters ($a/\xi'$, $r_i/\xi'$, $\alpha_{\uparrow}$ and $\alpha_{\downarrow}$) for $\abs{E} > E_g$ has been performed. For all used parameters, inside each local minigap four discrete states have been found for a nonmagnetic impurity, and either three or four impurity states in the case of a magnetic impurity. Two impurity states have never been found.

Speaking of impurity-induced sub-gap features, we have to mention the ordinary Shiba state. One may expect this state to exist when a magnetic impurity is located sufficiently far from the normal bubble. For a magnetic point impurity in a bulk superconductor the energy of the Shiba state is given by Eq. \eqref{eq:Rusinov} with $l=0$.
If $E_{S0} < E_g$, we identify one of our impurity state as the Shiba state. When $E_{S0} > E_g$ a discrete Shiba cannot appear because there are no local minigaps at distances $r>a$. Thus, the Shiba state becomes a resonance with a finite lifetime due to the possibility of quasiparticle tunneling between the magnetic impurity and the normal bubble.

\section{Conclusion}
\label{sec:Conclusion}

To sum up, we have analyzed the subgap density of states in a bulk superconductor with a normal spherical inclusion in the presence of a point impurity. We found that the impurity, whether magnetic or not, generally induces two discrete quasiparticle states with energies $E<E_g$, where $E_g$ is the minigap of the clean system. Additional discrete states may appear at higher energies. We have calculated the energies of the impurity states and their wave function for various positions of the impurity and scattering phases. Finally, we have demonstrated that the emergence of two discrete states induced by a point impurity should be a quite common feature of superconducting systems with a real and spatially inhomogeneous order parameter.

The obtained results are relevant in view of the problem of quasiparticle poisoning in superconducting devices. We have shown that when the order parameter is inhomogeneous, additional subgap states are induced by impurities, which should result in enhanced quasiparticle trapping due to the increased number of localized states for quasiparticles. This scenario might explain the enhanced quasiparticle recombination rate in the presence of nonmagnetic disorder \cite{Barends+2009PRB}. Also, the qualitative modification of the spectrum of Shiba states that we have found might be relevant for engineering magnetic chains hosting Majorana modes \cite{Nadj-Perge+2013PRB,Pientka+2013PRB,Vazifeh+PRL2013,Klinovaja+2013PRL,Braunecker+2013PRL}.

\begin{acknowledgments}
I am very grateful to A. S. Mel’nikov for helpful discussions and for thorough reading of this paper. The work has been supported by Russian Foundation for Basic Research Grant No. 18-42-520037  (Secs. \ref{sec:Impurity} -- general properties of impurity states, and Appendix \ref{app:ImpStates}) 
Russian Science Foundation Grant No. 17-12-01383 (Secs. \ref{sec:Basic} and \ref{sec:pure} and Appendix \ref{app:Andreev}) and No. 15-12-10020 (Sec. \ref{sec:Impurity} -- specific properties of bound states in the presence of a magnetic impurity), 
and Foundation for the advancement
of theoretical physics BASIS Grant No. 109  (Appendix \ref{app:realistic}).
\end{acknowledgments}

\appendix
\section{Quasiclassical Green functions}
\label{app:Andreev}

In this Appendix we obtain the quasiclassical Green functions for our system, transform the Green functions with coinciding arguments [Eqs. \eqref{eq:GER_qq} and \eqref{eq:FE_qq}] and derive Eqs. \eqref{eq:nu_in} and \eqref{eq:nu_out}.

\begin{figure}[htb]
	\centering
		\includegraphics[width = \linewidth]{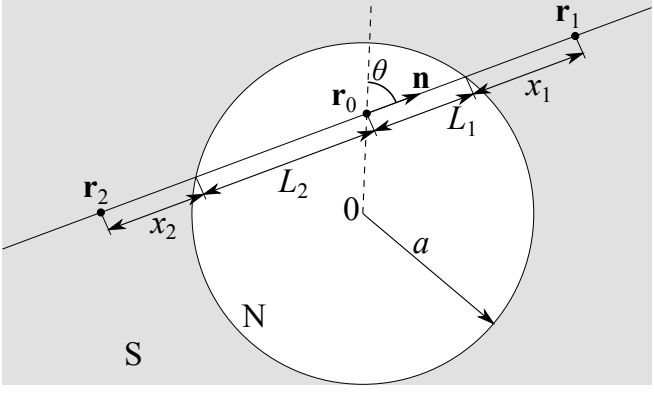}
	\caption{Explanatory image for Eqs. \eqref{eq:gE_in} - \eqref{eq:fE_incomming}.}
	\label{fig:Sphere}
\end{figure}

We start with the functions $g_E(\vec{r},\vec{n})$ and $f_E^{\dagger}(\vec{r},\vec{n})$, which satisfy the Eilenberger equations. For our system these equations have the same form as for a Josephson SNS junction, and their solution can be found in a textbook \cite{Svidzinskii_inhomogeneous}. Let us denote as $\theta$ the angle between $\vec{r}$ and $\vec{n}$. Then, the impact parameter of a classical (straight) trajectory with a direction vector $\vec{n}$ passing through the point $\vec{r}$ is $h = r \sin \theta$. If $h > a$, the trajectory does not cross the normal region, so that the Green functions are the same as in a bulk superconductor:
\begin{equation}
	g_E(\vec{r},\vec{n}) = -\frac{iE}{\sqrt{\Delta^2 - E^2}}, \qquad f_E^{\dagger}(\vec{r},\vec{n}) = -\frac{i\Delta}{\sqrt{\Delta^2 - E^2}}.
	\label{eq:gf_out}
\end{equation}
If $h < a$, the trajectory passes through the normal region. Such trajectory has three sections, where different expressions for the Green functions are valid. First, consider the section that lies inside the normal bubble, e.g., assume that $\vec{r} = \vec{r}_0$ -- see Fig. \ref{fig:Sphere}. Then the Green functions are
\begin{equation}
	g_E(\vec{r},\vec{n}) = i \cot \left( \frac{EL(r,\cos \theta)}{\hbar v_F} - \gamma(E) + i\eps \right),
	\label{eq:gE_in}
\end{equation}
\begin{equation}
	f_E^{\dagger}(\vec{r}_0,\vec{n}) = \frac{i e^{\frac{iE[L_1(r,\cos \theta)- L_2(r,\cos \theta)]}{\hbar v_F}}}{	\sin \left( \frac{EL(r,\cos \theta)}{\hbar v_F} - \gamma(E) + i\eps \right)},
	\label{eq:fE_in}
\end{equation}
where
\begin{equation}
	L_{1,2}(r,\cos \theta) = \mp r \cos \theta + \sqrt{a^2 - r^2 + r^2 \cos^2 \theta},
	\label{eq:L12}
\end{equation}
\begin{equation}
	L(r,\cos \theta) = 2\sqrt{a^2 - r^2 + r^2 \cos^2 \theta}.
	\label{eq:L}
\end{equation}
Next, consider $\vec{r} = \vec{r}_1$ [see Fig. \ref{fig:Sphere}]. Then
\begin{widetext}
\begin{equation}
	g_E(\vec{r},\vec{n}) = \left[ i \cot \left( \frac{EL(r,\cos \theta)}{\hbar v_F} - \gamma(E) + i\eps \right) + \frac{iE}{\sqrt{\Delta^2 - E^2}} \right] e^{-\frac{2\sqrt{\Delta^2- E^2}}{\hbar v_F} x_1(r,\cos \theta)} - \frac{iE}{\sqrt{\Delta^2 - E^2}},
	\label{eq:gE_outcomming}
\end{equation}
\begin{equation}
	f_E^{\dagger}(\vec{r},\vec{n}) = \left[ i \frac{e^{-\frac{iEL(r,\cos \theta)}{\hbar v_F}}}{\sin \left( \frac{EL(\vec{r},\cos \theta)}{\hbar v_F} - \gamma(E) + i\eps \right)} + \frac{i\Delta}{\sqrt{\Delta^2 - E^2}} \right] e^{-\frac{2\sqrt{\Delta^2- E^2}}{\hbar v_F} x_1(r,\cos \theta)} - \frac{i\Delta}{\sqrt{\Delta^2 - E^2}},
	\label{eq:fE_outcomming}
\end{equation}
where
\begin{equation}
	x_1(r,\cos \theta) = r \cos \theta - \sqrt{a^2 - r^2 + r^2 \cos^2 \theta}.
	\label{eq:x1}
\end{equation}
Finally, for $\vec{r} = \vec{r}_2$ [see Fig. \ref{fig:Sphere}] the Green functions are
\begin{equation}
	g_E(\vec{r},\vec{n}) = \left[ i \cot \left( \frac{EL(r,\cos \theta)}{\hbar v_F} - \gamma(E) + i\eps \right) + \frac{iE}{\sqrt{\Delta^2 - E^2}} \right] e^{-\frac{2\sqrt{\Delta^2- E^2}}{\hbar v_F} x_2(r,\cos \theta)} - \frac{iE}{\sqrt{\Delta^2 - E^2}},
	\label{eq:gE_incomming}
\end{equation}
\begin{equation}
	f_E^{\dagger}(\vec{r},\vec{n}) = \left[ i \frac{e^{\frac{iEL(r,\cos \theta)}{\hbar v_F}}}{\sin \left( \frac{EL(r,\cos \theta)}{\hbar v_F} - \gamma(E) + i\eps \right)} + \frac{i\Delta}{\sqrt{\Delta^2 - E^2}} \right] e^{-\frac{2\sqrt{\Delta^2- E^2}}{\hbar v_F} x_2(r,\cos \theta)} - \frac{i\Delta}{\sqrt{\Delta^2 - E^2}},
	\label{eq:fE_incomming}
\end{equation}
where
\begin{equation}
	x_2(r,\cos \theta) = -r\cos \theta - \sqrt{a^2 - r^2 + r^2 \cos^2 \theta}.
	\label{eq:x2}
\end{equation}

Using Eqs. \eqref{eq:gf_out} - \eqref{eq:x2}, we can transform Eqs. \eqref{eq:GER_qq} and \eqref{eq:FE_qq}. If we denote as $\theta$ the angle between $\vec{r}$ and $\vec{n}$, for $r < a$ we find that
\begin{equation}
	G_{ER}^{(0)}(\vec{r},\vec{r}) = \frac{imk_F}{4\pi \hbar^2} \int_0^{\pi} g_E(\vec{r},\vec{n}) \sin \theta d \theta = -\frac{mk_F}{2\pi \hbar^2} \int_0^1 \cot \left( \frac{E L(r,t)}{\hbar v_F} - \gamma(E) + i\eps \right) dt.
	\label{eq:GER_in}
\end{equation}
where we used the integration variable $t = \cos \theta$. Similarly, for $F_E^{\dagger (0)}(\vec{r},\vec{r})$ we have
\begin{equation}
		F_E^{\dagger(0)}(\vec{r},\vec{r}) = -\frac{mk_F}{2\pi \hbar^2} \int_0^1 \frac{\cos\left( \frac{2Er}{\hbar v_F} t \right)}{\sin \left( \frac{E L(r,t)}{\hbar v_F} - \gamma(E) + i\eps \right)} dt.
	\label{eq:FE_in}
\end{equation}
For $r >a$ we obtain
\begin{equation}
	G_{ER}^{(0)}(\vec{r},\vec{r}) = \frac{mk_F}{2\pi \hbar^2} \left\{ \frac{E}{\sqrt{\Delta^2 - E^2}}  - \!\!\!\!\!\!\! \int\limits_{\sqrt{1-a^2/r^2}}^1 \left[ \cot \left( \frac{EL(r,t)}{\hbar v_F} - \gamma(E) + i\eps \right) + \frac{E}{\sqrt{\Delta^2 - E^2}} \right] e^{-\frac{2\sqrt{\Delta^2- E^2}}{\hbar v_F} x_1(r,t)} dt \right\},
	\label{eq:GER_out}
\end{equation}
\begin{equation}
	F_E^{\dagger (0)}(\vec{r},\vec{r}) = \frac{mk_F}{2\pi \hbar^2} \left\{ \frac{\Delta}{\sqrt{\Delta^2 - E^2}}  - \!\!\!\!\!\!\! \int\limits_{\sqrt{1-a^2/r^2}}^1 \left[ \frac{\cos\left( \frac{EL(r,t)}{\hbar v_F} \right)}{\sin \left( \frac{EL(r,t)}{\hbar v_F} - \gamma(E) + i\eps \right)} + \frac{\Delta}{\sqrt{\Delta^2 - E^2}} \right] e^{-\frac{2\sqrt{\Delta^2- E^2}}{\hbar v_F} x_1(r,t)} dt \right\},
	\label{eq:FE_out}
\end{equation}
\end{widetext}
Generally, the integrals in Eqs. \eqref{eq:GER_in}, \eqref{eq:FE_in}, \eqref{eq:GER_out} and \eqref{eq:FE_out} cannot be evaluated analytically, however, for the local density of states we need only the imaginary part of $G_{ER}^{(0)}(\vec{r},\vec{r})$. To evaluate this quantity the following identity will be useful:
\begin{equation}
	\mathrm{Im} \left[ \cot(x + i\eps) \right] = -\pi \sum_{n = -\infty}^{+\infty} \delta(x - \pi n).
	\label{eq:cot_delta}
\end{equation}
Applying this to Eqs. \eqref{eq:GER_in} and \eqref{eq:GER_out}, we obtain Eqs. \eqref{eq:nu_in} and \eqref{eq:nu_out}.

The quasiclassical treatment of Green functions with noncoinciding coordinates is described in Ref. \cite{Gorkov+72JETP_eng}. There, the functions $G_E^{(0)}(\vec{r},\vec{r}')$ and $F_E^{\dagger (0)}(\vec{r},\vec{r}')$ are expressed in terms of retarded quasiclassical functions denoted as $g_{\pm}^R(\vec{r},\vec{r}')$ and $f_{\pm}^{\dagger R}(\vec{r},\vec{r}')$:
\begin{eqnarray}
	& G_E^{(0)}(\vec{r},\vec{r}') = \frac{m}{2\pi \hbar^2 \abs{\vec{r}- \vec{r}'}} \left[ g^R_+ (\vec{r},\vec{r}') e^{ik_F \abs{\vec{r} - \vec{r}'}} \right. & \nonumber \\
	& \left. + g^R_- (\vec{r},\vec{r}') e^{-ik_F \abs{\vec{r} - \vec{r}'}} \right]. &
	\label{eq:G_g_pm}
\end{eqnarray}
\begin{eqnarray}
	& F_E^{\dagger(0)}(\vec{r},\vec{r}') = \frac{m}{2\pi \hbar^2 \abs{\vec{r}- \vec{r}'}} \left[ f^{\dagger R}_+ (\vec{r},\vec{r}') e^{ik_F \abs{\vec{r} - \vec{r}'}} \right. & \nonumber \\
	& \left. + f^{\dagger R}_- (\vec{r},\vec{r}') e^{-ik_F \abs{\vec{r} - \vec{r}'}} \right]. &
	\label{eq:F_f_pm}
\end{eqnarray}
If one puts $\vec{r} = \vec{r}' + s\vec{n}$, the functions $g_{\pm}^R(\vec{r},\vec{r}')$ and $f_{\pm}^{\dagger R}(\vec{r},\vec{r}')$ will satisfy Andreev equations:
\begin{equation}
	\mp i\hbar v_F \frac{\partial g_{\pm}^R}{\partial s} - (E+i\eps) g_{\pm}^R + \Delta(\vec{r}' + s\vec{n}) f_{\pm}^{\dagger R} = 0,
	\label{eq:g_pm_Andreev}
\end{equation}
\begin{equation}
	\pm i\hbar v_F \frac{\partial f_{\pm}^{\dagger R}}{\partial s} + \Delta^*(\vec{r}' + s\vec{n}) g_{\pm}^R -  (E+i\eps) f_{\pm}^{\dagger R}  = 0.
	\label{eq:f_pm_Andreev}
\end{equation}
Here, $s$ is positive. The boundary conditions are
\begin{equation}
	g_{\pm}^R (\vec{r}' + s \vec{n},\vec{r}') \biggl|_{s=+0} = \frac{1}{2} [1 \pm g_E(\vec{r}',\pm \vec{n})]
	\label{eq:g_pm_bound}
\end{equation}
\begin{equation}
	f_{\pm}^{\dagger R} (\vec{r}' + s \vec{n},\vec{r}') \biggl|_{s=+0} = \pm \frac{1}{2} f_E^{\dagger}(\vec{r}',\pm \vec{n})
	\label{eq:f_pm_bound}
\end{equation}
From this it follows that
\begin{equation}
	g_+^R (\vec{r}' + s \vec{n},\vec{r}') \biggl|_{s=+0} + g_{-}^R (\vec{r}' - s \vec{n},\vec{r}') \biggl|_{s=+0} = 1,
	\label{eq:g_pm_bound1}
\end{equation}
\begin{equation}
	f_+^{\dagger R} (\vec{r}' + s \vec{n},\vec{r}') \biggl|_{s=+0} + f_-^{\dagger R} (\vec{r}' - s \vec{n},\vec{r}') \biggl|_{s=+0} = 0.
	\label{eq:f_pm_bound1}
\end{equation}
Let us define the functions $\tilde{g}_E$ and $\tilde{f}_E^{\dagger}$ as follows:
\begin{equation}
	\tilde{g}_E(\vec{r}',s,\vec{n}) = \left\{
	\begin{array}{ll}
	  g_+^R(\vec{r}' + s\vec{n},\vec{r}'), & s>0, \\
		-g_-^R(\vec{r}' + s\vec{n},\vec{r}'), & s<0,
	\end{array}
	\right.
	\label{eq:tg_def}
\end{equation}
\begin{equation}
	\tilde{f}^{\dagger}_E(\vec{r}',s,\vec{n}) = \left\{
	\begin{array}{ll}
	  f_+^{\dagger R}(\vec{r}' + s\vec{n},\vec{r}'), & s>0, \\
		-f_-^{\dagger R}(\vec{r}' + s\vec{n},\vec{r}'), & s<0.
	\end{array}
	\right.
	\label{eq:tf_def}
\end{equation}
One can see that these functions satisfy the inhomogeneous Andreev equations
\begin{widetext}
\begin{equation}
	-i\hbar v_F \frac{\partial \tilde{g}_E(\vec{r}',s,\vec{n})}{\partial s} - (E+i\eps) \tilde{g}_E(\vec{r}',s,\vec{n}) + \Delta(\vec{r}' + s\vec{n}) \tilde{f}_E^{\dagger}(\vec{r}',s,\vec{n}) = -i\hbar v_F \delta(s),
	\label{eq:g_Andreev}
\end{equation}
\begin{equation}
	i\hbar v_F \frac{\partial \tilde{f}_E^{\dagger}(\vec{r}',s,\vec{n})}{\partial s} + \Delta^*(\vec{r}' + s\vec{n}) \tilde{g}_E(\vec{r}',s,\vec{n}) -  (E+i\eps) \tilde{f}_E^{\dagger}(\vec{r}',s,\vec{n})  = 0,
	\label{eq:f_Andreev}
\end{equation}
\end{widetext}
and that Eqs. \eqref{eq:G_g_pm} and \eqref{eq:F_f_pm} can be written in the form \eqref{eq:GE(r,ri)} and \eqref{eq:FE(r,ri)}.

Let us list some properties of Eqs. \eqref{eq:g_Andreev} and \eqref{eq:f_Andreev}. We note that for $\abs{E} < \Delta$ the small imaginary term $i\eps$ is only relevant for a discrete set of energies, which correspond to subgap Andreev states. Here, we will consider only energies that are not in this set, so that $i\eps$ can be dropped. For such energies, one finds that
\begin{equation}
	\tilde{g}_{-E}(\vec{r}',-s,-\vec{n}) = - \tilde{g}_E(\vec{r}',s,\vec{n}), 
	\label{eq:invert_n_g}
\end{equation}
\begin{equation}
	\tilde{f}_{-E}^{\dagger}(\vec{r}',-s,-\vec{n}) = \tilde{f}_E^{\dagger}(\vec{r}',s,\vec{n}),
	\label{eq:invert_n_f}
\end{equation}
If $\Delta$ is real, for all energies (even when $i\eps$ is relevant) we have
\begin{equation}
	\tilde{g}_{-E}(\vec{r}',s,\vec{n}) = \tilde{g}_E^*(\vec{r}',s,\vec{n}),
	\label{eq:invert_E_g}
\end{equation}
\begin{equation}
	\tilde{f}_{-E}^{\dagger}(\vec{r}',s,\vec{n}) = -\tilde{f}_E^{\dagger*}(\vec{r}',s,\vec{n}).
	\label{eq:invert_E_f}
\end{equation}
Another property that follows from Eqs. \eqref{eq:g_Andreev} and \eqref{eq:f_Andreev} is
\begin{equation}
	\frac{\partial}{\partial s} \left( \abs{\tilde{g}_E}^2 - \abs{\tilde{f}_E^{\dagger}}^2 \right) = 0
	\label{eq:Andreev_conservation}
\end{equation}
for $s \neq 0$. Since at $s \to \pm \infty$ we have $\abs{\tilde{g}_E}^2 - \abs{\tilde{f}_E^{\dagger}}^2 \to 0$, it follows that
\begin{equation}
	\abs{\tilde{g}_E} =  \abs{\tilde{f}_E^{\dagger}}
	\label{eq:abs_g=abs_f}
\end{equation}
for all $s$.

\begin{figure}[htb]
	\centering
		\includegraphics[width = \linewidth]{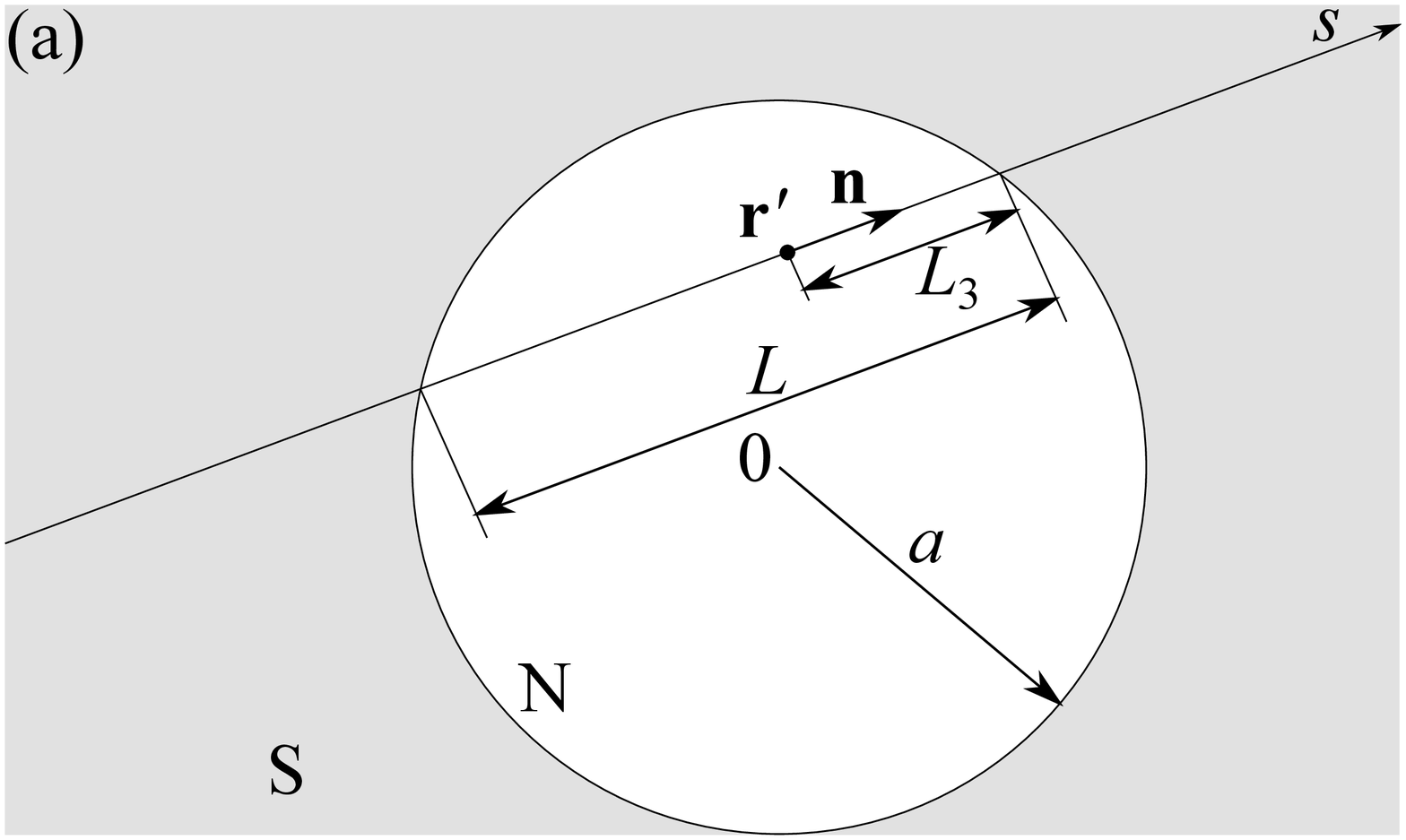}
		\includegraphics[width = \linewidth]{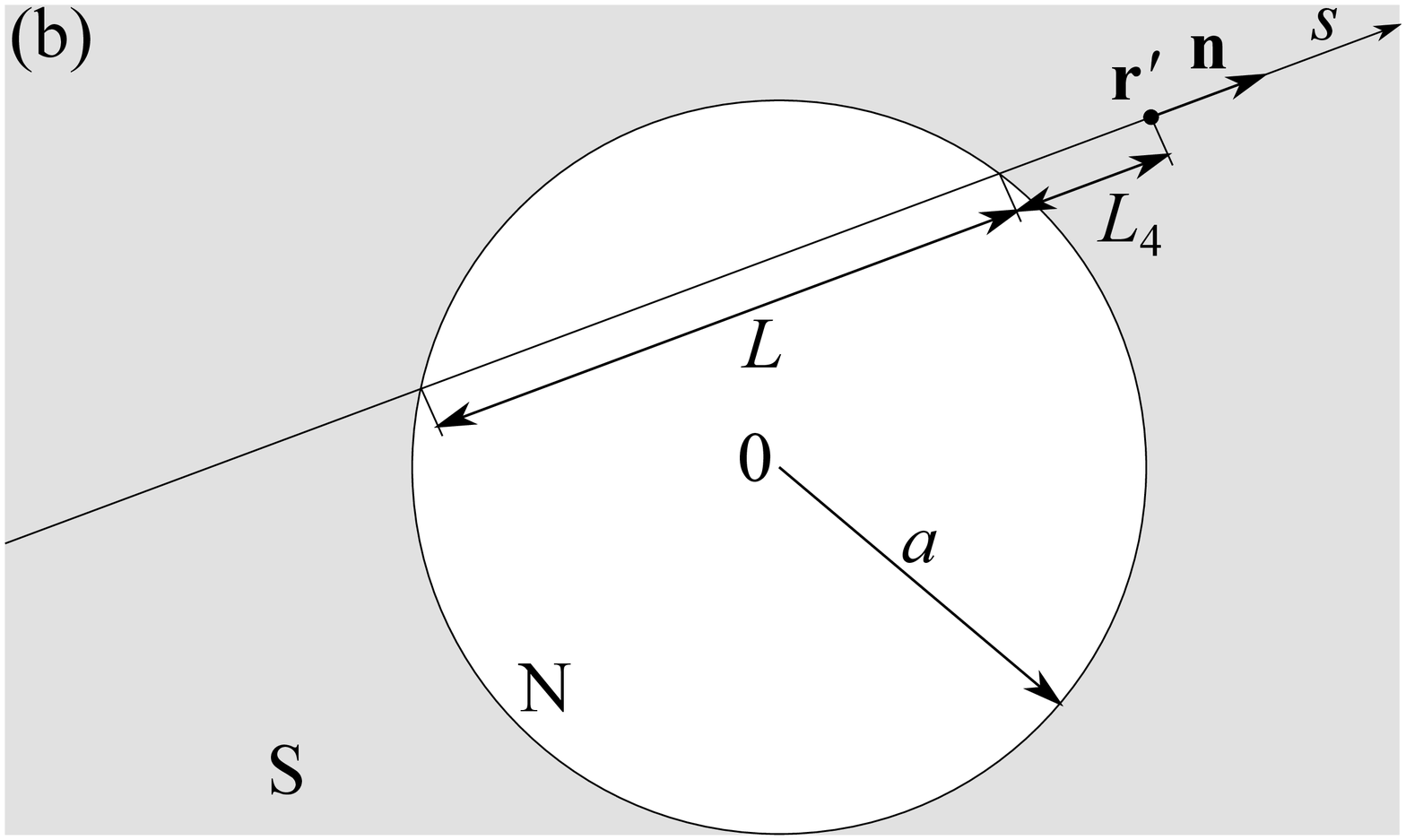}
		\includegraphics[width = \linewidth]{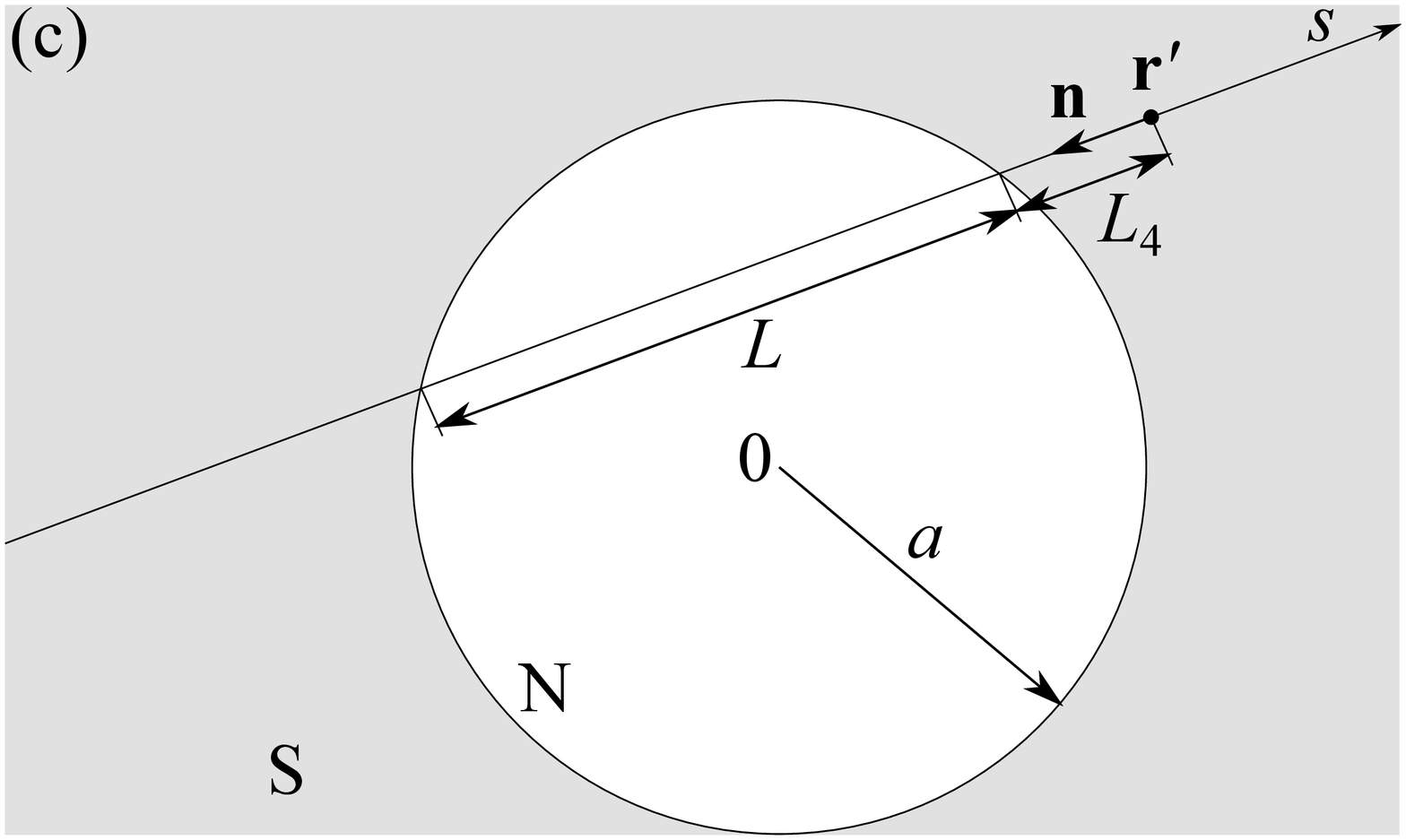}
	\caption{Explanatory images for the solutions of Eqs. \eqref{eq:g_Andreev} and \eqref{eq:f_Andreev}.}
	\label{fig:ri}
\end{figure}

Now we write down the solutions of Eqs. \eqref{eq:g_Andreev} and \eqref{eq:f_Andreev} for our system without impurities. Note that these equations are linear and have piecewise constant coefficients, so solving them is straightforward. We need to consider several cases depending on the position of $\vec{r}'$ and the direction of $\vec{n}$. First, let $\vec{r}'$ be inside the normal bubble ($r' < a$) -- see Fig. \ref{fig:ri}a. For $0<s< L_3$ we have then
\begin{equation}
	\tilde{g}_E = \frac{e^{\frac{iEs}{\hbar v_F}}}{1 - e^{\frac{2iEL}{\hbar v_F} - 2i\gamma}},
	\label{eq:gE_r1<a_0<s<L1}
\end{equation}
\begin{equation}
	\tilde{f}_E^{\dagger} = \frac{e^{\frac{2iEL_3}{\hbar v_F} - i\gamma -\frac{iEs}{\hbar v_F}}}{1 - e^{\frac{2iEL}{\hbar v_F} - 2i\gamma}},
	\label{eq:fE_r1<a_0<s<L1}
\end{equation}
and for $s>L_3$
\begin{equation}
	\tilde{g}_E = \frac{e^{\frac{iEL_3}{\hbar v_F}-\frac{\sqrt{\Delta^2 - E^2}}{\hbar v_F}(s-L_3)}}{1 - e^{\frac{2iEL}{\hbar v_F} - 2i\gamma}},
	\label{eq:gE_r1<a_x>L1}
\end{equation}
\begin{equation}
	\tilde{f}_E^{\dagger} = \frac{e^{\frac{iEL_3}{\hbar v_F} - i\gamma - \frac{\sqrt{\Delta^2 - E^2}}{\hbar v_F}(s-L_3)}}{1 - e^{\frac{2iEL}{\hbar v_F} - 2i\gamma}}.
	\label{eq:fE_r1<a_x>L1}
\end{equation}
We will not write down the functions $\tilde{g}_E$ and $\tilde{f}_E^{\dagger}$ for $s<0$, since they can be obtained using Eqs. \eqref{eq:invert_n_g} and \eqref{eq:invert_n_f}.

Next, consider $r' > a$ and assume that the trajectory parametrized by $s$ crosses the normal region. Let $(\vec{r}' \vec{n})$ be positive, as in Fig. \ref{fig:ri}b. Then the Green functions for $s>0$ are
\begin{equation}
	\tilde{g}_E = \left[ \frac{e^{-\frac{2\sqrt{\Delta^2 - E^2} L_4}{\hbar v_F}}\left( e^{-\frac{2iEL}{\hbar v_F}} -1 \right)}{1 - e^{-\frac{2iEL}{\hbar v_F} + 2i\gamma}} + 1 \right] \frac{e^{- \frac{\sqrt{\Delta^2 - E^2}s}{\hbar v_F}}}{1 - e^{-2i\gamma}},
	\label{eq:gE_r1>a_out_s>0}
\end{equation}
\begin{equation}
	\tilde{f}^{\dagger}_E = \left[ \frac{e^{-\frac{2\sqrt{\Delta^2 - E^2} L_4}{\hbar v_F}}\left( e^{-\frac{2iEL}{\hbar v_F}} -1 \right)}{1 - e^{-\frac{2iEL}{\hbar v_F} + 2i\gamma}} +1 \right] \frac{e^{-i\gamma - \frac{\sqrt{\Delta^2 - E^2}s}{\hbar v_F}}}{1 - e^{-2i\gamma}}.
	\label{eq:fE_r1>a_out_s>0}
\end{equation}

Now let the vector $\vec{n}$ point in the opposite direction -- see Fig. \ref{fig:ri}c. The Green functions are
\begin{eqnarray}
	& \tilde{g}_E = \frac{1}{1 - e^{-2i\gamma}} \left[ \frac{e^{-\frac{2\sqrt{\Delta^2 - E^2}L_4}{\hbar v_F} - 2i\gamma} \left(e^{\frac{2iEL}{\hbar v_F}} -1 \right)}{1 - e^{-2i\gamma + \frac{2iEL}{\hbar v_F}}} e^{\frac{\sqrt{\Delta^2 -E^2}s}{\hbar v_F}} \right. & \nonumber \\
	& \left. + e^{-\frac{\sqrt{\Delta^2 - E^2}s}{\hbar v_F}} \right], &
	\label{eq:gE_r1>a_in_0<s<L1}
\end{eqnarray}
\begin{eqnarray}
	& \tilde{f}^{\dagger}_E = \frac{e^{-i\gamma}}{1 - e^{-2i\gamma}} \left[ \frac{e^{-\frac{2\sqrt{\Delta^2 - E^2}L_4}{\hbar v_F}} \left(e^{\frac{2iEL}{\hbar v_F}} -1 \right)}{1 - e^{-2i\gamma + \frac{2iEL}{\hbar v_F}}} e^{\frac{\sqrt{\Delta^2 -E^2}s}{\hbar v_F}} \right. & \nonumber \\
	& \left. + e^{-\frac{\sqrt{\Delta^2 -E^2}s}{\hbar v_F}} \right] &
	\label{eq:fE_r1>a_in_0<s<L1}
\end{eqnarray}
for $0<s<L_4$,
\begin{equation}
	\tilde{g}_E = \frac{e^{\frac{iE(s-L_4)}{\hbar v_F} - \frac{\sqrt{\Delta^2 -E^2}L_4}{\hbar v_F}}}{1 - e^{-2i\gamma + \frac{2iEL}{\hbar v_F}}},
	\label{eq:gE_r1>a_in_L1<s<L+L1}
\end{equation}
\begin{equation}
	\tilde{f}_E^{\dagger} = \frac{e^{\frac{iE}{\hbar v_F}(2L + L_4 - s) - i\gamma - \frac{\sqrt{\Delta^2 -E^2}L_4}{\hbar v_F}}}{1 - e^{-2i\gamma + \frac{2iEL}{\hbar v_F}}}
	\label{eq:fE_r1>a_in_L1<s<L+L1}
\end{equation}
for $L_1<s<L+L_4$, and
\begin{equation}
	\tilde{g}_E = \frac{e^{\frac{\sqrt{\Delta^2 -E^2}(L-s)}{\hbar v_F} + \frac{iEL}{\hbar v_F}}}{1 - e^{-2i\gamma + \frac{2iEL}{\hbar v_F}}},
	\label{eq:gE_r1>a_in_s>L+L1}
\end{equation}
\begin{equation}
	\tilde{f}^{\dagger}_E = \frac{e^{\frac{\sqrt{\Delta^2 -E^2}(L-s)}{\hbar v_F} + \frac{iEL}{\hbar v_F}- i\gamma}}{1 - e^{-2i\gamma + \frac{2iEL}{\hbar v_F}}}
	\label{eq:fE_r1>a_in_s>L+L1}
\end{equation}
for $s>L+L_4$.

Finally, there are trajectories that do not pass through the normal region. To obtain the Green functions on such trajectories, one may simply substitute $L = 0$ into Eqs. \eqref{eq:gE_r1>a_in_s>L+L1} and \eqref{eq:fE_r1>a_in_s>L+L1}:
\begin{equation}
	\tilde{g}_E = \frac{e^{-\frac{\sqrt{\Delta^2 -E^2}}{\hbar v_F}s }}{1 - e^{-2i\gamma}},
	\label{eq:gE_bulk}
\end{equation}
\begin{equation}
	\tilde{f}^{\dagger}_E = = \frac{e^{-\frac{\sqrt{\Delta^2 -E^2}}{\hbar v_F}s -i\gamma}}{1 - e^{-2i\gamma}}
	\label{eq:fE_bulk}
\end{equation}
for $s > 0$.

\section{Impurity-induced states: proof of existence and wave functions}
\label{app:ImpStates}

In this Appendix we will prove that ${\cal D}_{\uparrow}(E)$ [Eq. \eqref{eq:D_up_simplified}] has two roots at $\abs{E}<E_g$ and we will find the wave functions of impurity-induced states. Also, impurity states inside local minigaps will be briefly considered.

%%%%%%%%%% Proof of Eq. eq:F'<G'%%%%%%%%%%%%

We start by deriving a general analytical property of the Green functions with coinciding arguments. We will make use of the following relations \cite{Kopnin-book}:
\begin{equation}
	G_E^{(0)}(\vec{r},\vec{r}') = \sum_n \frac{u_n^{(0)}(\vec{r}) u^{(0)*}_n(\vec{r}')}{E_n - E - i\eps},
	\label{eq:G_BdG}
\end{equation}
\begin{equation}
	F_E^{\dagger (0)}(\vec{r},\vec{r}') = \sum_n \frac{v^{(0)}_n(\vec{r}) u^{(0)*}_n(\vec{r}')}{E_n - E - i\eps},
	\label{eq:F_BdG}
\end{equation}
where $(u_n^{(0)}(\vec{r}),v_n^{(0)}(\vec{r}))^T$ are the quasiparticle wave functions of the clean system and $E_n$ are the quasiparticle energies. Let us differentiate Eqs. \eqref{eq:G_BdG} and \eqref{eq:F_BdG} with respect to energy, taking $\vec{r} = \vec{r}'$ and $E \neq E_n$:
\begin{equation}
	G_{ER}^{(0)\prime}(\vec{r},\vec{r}) = \sum_{n>0} \left[ \frac{\abs{u_n^{(0)}(\vec{r})}^2}{(E_n - E)^2} + \frac{\abs{v_n^{(0)}(\vec{r})}^2}{(E_n + E)^2} \right],
	\label{eq:G'1}
\end{equation}
\begin{eqnarray}
	& F_E^{\dagger (0) \prime}(\vec{r},\vec{r})  =  \sum\limits_{n>0} v^{(0)}_n(\vec{r}) u^{(0)*}_n(\vec{r}) & \nonumber \\
	& \times \left[\frac{1}{(E_n - E)^2} -  \frac{1}{(E_n + E)^2} \right], &
	\label{eq:F'1}
\end{eqnarray}
where summation goes over states with positive energies, and we have used the fact that the wave functions of the states with negative energies $-E_n$ have the form $(v_n^{(0)*}(\vec{r}),-u_n^{(0)*}(\vec{r}))^T$. Within the quasiclassical approximation also $G_{ER}^{(0)}(\vec{r},\vec{r}) = - G_{-ER}^{(0)}(\vec{r},\vec{r})$, which follows from Appendix \ref{app:Andreev} (in particular, from Eq. \eqref{eq:invert_n_g}). Hence,
\begin{eqnarray}
	& G_{ER}^{(0)\prime}(\vec{r},\vec{r}) \approx \frac{1}{2} \left[ G_{ER}^{(0)\prime}(\vec{r},\vec{r}) + G_{-ER}^{(0)\prime}(\vec{r},\vec{r}) \right]  = & \nonumber \\
	& = \! \sum\limits_{n>0} \!\! \frac{\abs{u_n^{(0)}(\vec{r})}^2 + \abs{v_n^{(0)}(\vec{r})}^2}{2} \left[ \frac{1}{(E_n - E)^2} + \frac{1}{(E_n + E)^2} \right] \!\!. &
	\label{eq:G'2}
\end{eqnarray}
Since
\[ \abs{v^{(0)}_n(\vec{r}) u^{(0)*}_n(\vec{r})} \leq \frac{\abs{u_n^{(0)}(\vec{r})}^2 + \abs{v_n^{(0)}(\vec{r})}^2}{2} \]
and 
\[ \frac{1}{(E_n - E)^2} + \frac{1}{(E_n + E)^2} > \abs{ \frac{1}{(E_n - E)^2} - \frac{1}{(E_n + E)^2}}, \]
one can see from Eqs. \eqref{eq:F'1} and \eqref{eq:G'2} that
\begin{equation}
	\abs{F_E^{\dagger(0)\prime}(\vec{r}_i , \vec{r}_i)} < G_{ER}^{(0)\prime}(\vec{r}_i,\vec{r}_i).
	\label{eq:F'<G'}
\end{equation}
This relation is valid for all $E \in (-E_g,E_g)$.
%%%%%%%%%%%%%%%%%%%%%%%%%%%%%%%%%%%%%%%%%%%%%%%%%%%%%%%%%%%%%%%%%End of proof

Now we write ${\cal D}_{\uparrow}(E)$ in the following form:
\begin{equation}
	{\cal D}_{\uparrow}(E) = - {\cal D}_{\uparrow+}(E) {\cal D}_{\uparrow-}(E),
	\label{eq:D_factors}
\end{equation}
where
\begin{eqnarray}
	& {\cal D}_{\uparrow \pm}(E) = G_{ER}(\vec{r}_i,\vec{r}_i) + \frac{mk_F}{4\pi \hbar^2}(\cot \alpha_{\downarrow} - \cot \alpha_{\uparrow}) & \nonumber \\
	& \hspace{-0.5cm} \pm \sqrt{\left(\frac{mk_F}{4\pi \hbar^2} \right)^2 (\cot \alpha_{\downarrow} + \cot \alpha_{\uparrow})^2 + \abs{F_E^{\dagger}(\vec{r}_i,\vec{r}_i)}^2}. &
	\label{eq:D+-}
\end{eqnarray}
By direct differentiation and using Eq. \eqref{eq:F'<G'} it can be proven that ${\cal D}_{\uparrow+}(E)$ and ${\cal D}_{\uparrow-}(E)$ are strictly monotonic functions of energy.

Let us evaluate $G_{ER}^{(0)}(\vec{r}_i,\vec{r}_i)$ and $F_E^{\dagger (0)}(\vec{r}_i,\vec{r}_i)$ at $E \to E_g - 0$. For $r_i <a$ we use Eqs. \eqref{eq:GER_in} and \eqref{eq:FE_in}. When $r_i \neq 0$, the main contribution to the integrals comes from $t \approx 1$. Using this, we obtain
\begin{eqnarray}
  & G_{ER}^{(0)}(\vec{r}_i,\vec{r}_i) \approx \frac{m k_F}{2\pi \hbar^2} \int\limits_0^1 \frac{dt}{\gamma(E) - \frac{E L(r_i,t)}{\hbar v_F}} & \nonumber \\
	&  \approx \frac{m k_F}{2\pi \hbar^2} \int\limits_0^1 \frac{dt}{\gamma(E) - \frac{2Ea}{\hbar v_F}\left[ 1 - \frac{r_i^2}{a^2} (1-t) \right]} & \nonumber \\
	& \approx -\frac{k_F^2 a}{4\pi E r_i^2} \ln \left( \gamma(E) - \frac{2Ea}{\hbar v_F} \right). &
	\label{eq:G_Eg_r<a}
\end{eqnarray}
Here, in the denominator only the first two terms of its Taylor series in powers of $1-t$ have been retained. For $F_E^{\dagger (0)}(\vec{r}_i,\vec{r}_i)$ one finds that
\begin{equation}
	F_E^{\dagger (0)} (\vec{r}_i,\vec{r}_i) \approx G_{ER}^{(0)}(\vec{r}_i,\vec{r}_i) \cos \left( \frac{2E_g r_i}{\hbar v_F} \right).
	\label{eq:F_Eg_r<a}
\end{equation}
For $r_i >a$ from Eqs. \eqref{eq:GER_out} and \eqref{eq:FE_out} we find that
\begin{equation}
	  G_{ER}^{(0)}(\vec{r}_i,\vec{r}_i) \! \approx \! -\frac{k_F^2 a}{4\pi E r_i^2} \ln \!\! \left( \gamma(E) - \frac{2Ea}{\hbar v_F} \right) \! e^{-\frac{2\sqrt{\Delta^2- E^2}}{\hbar v_F}(r_i-a)} \!,
  \label{eq:G_Eg_r>a}
\end{equation}
\begin{equation}
	F_E^{\dagger (0)} (\vec{r}_i,\vec{r}_i) \approx G_{ER}^{(0)}(\vec{r}_i,\vec{r}_i) \cos \left( \frac{2E_g a}{\hbar v_F} \right).
	\label{eq:F_Eg_r>a}
\end{equation}
Since $2E_g a/(\hbar v_F)<\pi/2$, one can see that ${\cal D}_{\uparrow \pm}(E) \to +\infty$ as $E \to E_g -0$. Acording to Eqs. \eqref{eq:G_-ER} and \eqref{eq:F_-E}, for $\abs{E}<E_g$ $G_{ER}^{(0)}(\vec{r}_i,\vec{r}_i)$ is odd in $E$ and $F_E^{\dagger(0)}(\vec{r}_i,\vec{r}_i)$ is even in $E$, so that ${\cal D}_{\uparrow \pm}(E) \to -\infty$ as $E \to -E_g +0$. 
Hence each of the functions ${\cal D}_{\uparrow+}(E)$ and ${\cal D}_{\uparrow-}(E)$ turns to zero at a single value of $E$. This completes the proof of the statement that ${\cal D}_{\uparrow}(E) = 0$ has exactly two roots when $E \in (-E_g,E_g)$.

We conclude this Appendix by writing down the wave function of a spin-up impurity state. Let us take the larger root of ${\cal D}_{\uparrow}(E) = 0$, which we denote as $E = E_{1\uparrow}$. This is also the root of ${\cal D}_{\uparrow-}(E) = 0$, since ${\cal D}_{\uparrow-}(E) < {\cal D}_{\uparrow+}(E)$. Hence,
\begin{equation}
	{\cal D}_{\uparrow}'(E_{1\uparrow}) = -{\cal D}_{\uparrow-}'(E_{1\uparrow}) {\cal D}_{\uparrow+}(E_{1\uparrow}) <0.
	\label{eq:D'(E1)}
\end{equation}

Equations \eqref{eq:GE_+impurity} and \eqref{eq:GE(1)} yield
\begin{equation}
	G_{E_{1\uparrow} \uparrow \uparrow}(\vec{r},\vec{r}') \approx \frac{i}{\eps} u_{1\uparrow}(\vec{r}) u_{1\uparrow}^*(\vec{r}'),
	\label{eq:G(E_1uparrow)}
\end{equation}  
where
\begin{equation}
	u_{1\uparrow}(\vec{r}) = A_{\uparrow} G_{E_{1\uparrow}}^{(0)}(\vec{r},\vec{r}_i) - B_{\uparrow} F_{-E_{1\uparrow}}^{\dagger(0)*}(\vec{r},\vec{r}_i),
	\label{eq:u_imp}
\end{equation}
and the real quantities $A_{\uparrow}$ and $B_{\uparrow}$ are given by
\begin{equation}
	A_{\uparrow} = \sqrt{-{\cal D}_{\uparrow}'(E_{1\uparrow})^{-1} \left( G_{E_{1\uparrow}R}^{(0)}(\vec{r}_i,\vec{r}_i) + \frac{mk_F}{2\pi \hbar^2} \cot \alpha_{\downarrow} \right)},
	\label{eq:A_up}
\end{equation}
\begin{equation}
	B_{\uparrow} = \sqrt{-{\cal D}_{\uparrow}'(E_{1\uparrow})^{-1} \left( G_{E_{1\uparrow}R}^{(0)}(\vec{r}_i,\vec{r}_i) - \frac{mk_F}{2\pi \hbar^2} \cot \alpha_{\uparrow} \right)}.
	\label{eq:B_up}
\end{equation}
Here, we have used that
\begin{equation}
	G_E^{(0)}(\vec{r}',\vec{r}) = G_E^{(0)*}(\vec{r},\vec{r}'), \quad F_E^{\dagger (0)}(\vec{r}',\vec{r}) = F_{-E}^{\dagger (0)}(\vec{r},\vec{r}'),
	\label{eq:rr'_swap}
\end{equation}
which follows from Eqs. \eqref{eq:G_BdG} and \eqref{eq:F_BdG}. A generalized form of Eq. \eqref{eq:G_BdG} is also applicable to $G_{E_{1\uparrow} \uparrow \uparrow}(\vec{r},\vec{r}')$, from which we conclude that $u_{1\uparrow}(\vec{r})$ is the electron component of the wave function of the impurity state. The hole component of the wave function is then
\begin{equation}
	v_{1\uparrow}(\vec{r}) = A_{\uparrow} F_{E_{1\uparrow}}^{\dagger(0)}(\vec{r},\vec{r}_i) + B_{\uparrow} G_{-E_{1\uparrow}}^{(0)*}(\vec{r},\vec{r}_i).
	\label{eq:v_imp}
\end{equation}
This can be checked by substituting $(u_{1\uparrow}(\vec{r}),v_{1\uparrow}(\vec{r}))^T$ into the BdG equations. The wave function of the second quasiparticle state can be obtained in a similar way, so we do not write down here the corresponding expressions.

Finally, we will prove that discrete impurity states appear also inside local minigaps (for $E>E_g$). Let us take a normal bubble with radius $a > \xi'$. A local minigap for $r = r_i$ is an energy interval $E \in (E_A^{(n-1)}(r_i),E_A^{(n)}(0))$ with $n = 1.. \floor{a/\xi'}$, where $E_A^{(n)}(r)$ is the solution of the following equation:
\begin{equation}
	\frac{2E_A^{(n)}\sqrt{a^2 - r^2}}{\hbar v_F} - \gamma\left(E_A^{(n)} \right) = \pi n.
	\label{eq:E_A^(n)(r)}
\end{equation}
Of course, the energy interval $(E_A^{(n-1)}(r_i),E_A^{(n)}(0))$ exists only if $E_A^{(n-1)}(r_i) < E_A^{(n)}(0)$, which is satisfied when $r_i$ is sufficiently small:
\begin{equation}
	r_i < \sqrt{\frac{\pi \hbar v_F}{E_A^{(n)}(0)}} \sqrt{a - \frac{\pi \hbar v_F}{4 E_A^{(n)}(0)}}.
	\label{eq:ri_small}
\end{equation}
For $E \in (E_A^{(n-1)}(r_i),E_A^{(n)}(0))$ it can be seen from Eqs. \eqref{eq:GER_in} and \eqref{eq:FE_in} that the Green functions with coinciding arguments are real. Moreover, they diverge when the energy approaches the boundaries of the given interval. For $E \to E_A^{(n-1)}(r_i) + 0$ the main contribution to the integrals in Eqs. \eqref{eq:GER_in} and \eqref{eq:FE_in} comes from $t \approx 0$. Then, acting in the same way as when deriving Eq. \eqref{eq:G_Eg_r<a}, we find that
\begin{eqnarray}
  & G_{ER}^{(0)}(\vec{r}_i,\vec{r}_i) \approx - \frac{mk_F}{4\hbar^2} & \nonumber \\
	& \hspace{-0.7cm} \times \left[ \left( \frac{2E \sqrt{a^2 - r_i^2}}{\hbar v_F} - \gamma(E) - \pi (n-1) \right) \frac{E_A^{(n-1)}(r_i) r_i^2}{\hbar v_F \sqrt{a^2 - r_i^2}} \right]^{-1/2}  \!\!\!\!\!\!\!\!\!\!\!\!\! , & \label{eq:GER_EA(r)} \\
	& F_E^{\dagger (0)}(\vec{r}_i,\vec{r}_i)  \approx (-1)^{n-1} G_{ER}^{(0)}(\vec{r}_i,\vec{r}_i). &
	\label{eq:FE_EA(r)}
\end{eqnarray}
Thus, we have an inverse-square-root divergence. Similarly, for $E \to E_A^{(n)}(0) - 0$ we obtain
\begin{equation}
	G_{ER}^{(0)}(\vec{r}_i,\vec{r}_i) \approx -\frac{k_F^2 a}{4\pi E_A^{(n)}(0) r_i^2} \ln \left( \pi n + \gamma(E) - \frac{2Ea}{\hbar v_F} \right),
	\label{eq:GER_EA(0)}
\end{equation}
\begin{equation}
	F_E^{\dagger (0)}(\vec{r}_i,\vec{r}_i) \approx (-1)^n \cos \left( \frac{2E_A^{(n)}(0) r_i}{\hbar v_F} \right) G_{ER}^{(0)}(\vec{r}_i,\vec{r}_i).
	\label{eq:FE_EA(0)}
\end{equation}
This means that ${\cal D}_{\uparrow \pm}(E) \to +\infty$ [Eq. \eqref{eq:D+-}] when $E \to E_A^{(n)}(0) - 0$, unless
\begin{equation}
	\cos \left( \frac{2E_A^{(n)}(0) r_i}{\hbar v_F} \right) = \pm 1.
	\label{eq:cos_unlikely}
\end{equation}
In addition, one can see from Eqs. \eqref{eq:GER_EA(r)} and \eqref{eq:FE_EA(r)} that ${\cal D}_{\uparrow-}(E) \to -\infty$ when $E \to  E_A^{(n-1)}(r_i)+0$. According to considerations above, this means that the equation ${\cal D}_{\uparrow}(E) = 0$ has one or two roots for $E \in (E_A^{(n-1)}(r_i),E_A^{(n)}(0))$. The same arguments apply to the energy interval $E \in (-E_A^{(n)}(0),-E_A^{(n-1)}(r_i))$ (which can be proved using that $G_{ER}^{(0)}(\vec{r}_i,\vec{r}_i)$ and $F_E^{\dagger (0)}(\vec{r}_i,\vec{r}_i)$ are odd and even in $E$, respectively). Hence, there are no less than one and no more than two impurity states with each spin projection in the given energy interval.

\section{Sufficient condition for the existence of two impurity states}
\label{app:realistic}

In this appendix we derive a quite general sufficient condition for the existence of two impurity levels inside a gap of an inhomogeneous superconductor with a point impurity.
 
According to Appendix \ref{app:ImpStates}, the roots of the equations ${\cal D}_{\uparrow+}(E)= 0$ and ${\cal D}_{\uparrow-}(E)= 0$ [see Eq. \eqref{eq:D+-}] yield the energies of the discrete states induced in a superconducting system by a point impurity with scattering phases $\alpha_{\uparrow} \neq 0$ and $\alpha_{\downarrow} \neq 0$. If the system without impurities can be described within the quasiclassical approximation, ${\cal D}_{\uparrow+}(E)$ and ${\cal D}_{\uparrow-}(E)$ are monotonically increasing functions of energy. Since ${\cal D}_{\uparrow+}(E) \geq {\cal D}_{\uparrow-}(E)$, the {\it necessary and sufficient} conditions for the existence of two impurity states then read
\begin{equation}
	\lim_{E \to E_g-0} {\cal D}_{\uparrow-}(E) > 0,
	\label{eq:2states_+}
\end{equation}
\begin{equation}
	\lim_{E \to -E_g+0} {\cal D}_{\uparrow+}(E) < 0.
	\label{eq:2states_-}
\end{equation}
We will describe a class of systems for which these conditions are satisfied.

Consider a clean superconducting system with an inhomogeneous real order parameter $\Delta(\vec{r})$, such that $\Delta(\vec{r}) \to \Delta_{\infty}>0$ when $r \to \infty$. We will assume that the energy spectrum of this system has a finite gap $E_g \leq \Delta_{\infty}$ and that the system can be described within the quasiclassical approximation. For the quasiclassical Green functions a useful Riccati parametrization exists \cite{Schopohl98}:
\begin{equation}
	g_E(\vec{r},\vec{n}) = \frac{1 - a_E(\vec{r},\vec{n}) b_E(\vec{r},\vec{n})}{1 + a_E(\vec{r},\vec{n}) b_E(\vec{r},\vec{n})},
	\label{eq:g_Riccati}
\end{equation}
\begin{equation}
	f_E^{\dagger}(\vec{r}_i,\vec{n}) = - \frac{2ib_E(\vec{r},\vec{n})}{1 + a_E(\vec{r},\vec{n}) b_E(\vec{r},\vec{n})}.
	\label{eq:f+_Riccati}
\end{equation}
\begin{equation}
	f_E(\vec{r}_i,\vec{n}) = - \frac{2ia_E(\vec{r},\vec{n})}{1 + a_E(\vec{r},\vec{n}) b_E(\vec{r},\vec{n})}.
	\label{eq:f_Riccati}
\end{equation}
In our case the Riccati amplitudes $a_E(\vec{r},\vec{n})$ and $b_E(\vec{r},\vec{n})$ satisfy the following equations:
\begin{equation}
	\hbar v_F \vec{n} \nabla a_E + [\Delta(\vec{r}) a_E -2i(E +i\eps)] a_E - \Delta(\vec{r}) = 0, \label{eq:a_E} 
\end{equation}
\begin{equation}
	\hbar v_F \vec{n} \nabla b_E - [\Delta(\vec{r}) b_E -2i(E +i\eps)] b_E + \Delta(\vec{r}) = 0. \label{eq:b_h}
\end{equation}
These equations should be solved on classical trajectories, which can be parameterized by a variable $s$: $\vec{r} = \vec{r}_0 + s\vec{n}$, where $\vec{r}_0$ is some point on the trajectory. Then, the boundary conditions read
\begin{eqnarray}
	& a_E \biggl|_{s\to -\infty} = b_E \biggl|_{s\to +\infty} & \nonumber \\
	& = \frac{\Delta_{\infty}}{-i(E + i\eps) + \sqrt{\Delta_{\infty}^2 - (E+i\eps)^2}}. &
	\label{eq:a_bound}
\end{eqnarray}
For our purpose, the following parametrization is more convenient:
\begin{equation}
	a_E(\vec{r},\vec{n}) = ie^{i \alpha_E(\vec{r},\vec{n})}, \quad b_E(\vec{r},\vec{n}) = ie^{i \beta_E(\vec{r},\vec{n})}.
	\label{eq:alpha_beta}
\end{equation}
The functions $\alpha_E(\vec{r},\vec{n})$ and $\beta_E(\vec{r},\vec{n})$ satisfy the differential equations
\begin{equation}
	\hbar v_F \frac{\partial \alpha_E}{\partial s} = 2(E + i\eps) - 2\Delta(\vec{r}) \cos \alpha_E,
	\label{eq:alpha_diff}
\end{equation}
\begin{equation}
	\hbar v_F \frac{\partial \beta_E}{\partial s} = -2(E + i\eps) + 2\Delta(\vec{r}) \cos \beta_E,
	\label{eq:beta_diff}
\end{equation}
with the boundary conditions
\begin{equation}
	\alpha_E \biggl|_{s \to -\infty} = \beta_E \biggl|_{s \to +\infty} = -\arccos \left( \frac{E + i\eps}{\Delta_{\infty}} \right).
	\label{eq:ab_bound}
\end{equation}
The quasiclassial Green functions are expressed in terms of $\alpha_E$ and $\beta_E$ as follows:
\begin{equation}
	g_E = i \cot \left( \frac{\alpha_E + \beta_E}{2} \right),
	\label{eq:g<-ab}
\end{equation}
\begin{equation}
	f_E^{\dagger} = \frac{2e^{i\beta_E}}{1 - e^{i\alpha_E + i\beta_E}},
	\label{eq:f+<-ab}
\end{equation}
\begin{equation}
	f_E = \frac{2e^{i\alpha_E}}{1 - e^{i\alpha_E + i\beta_E}}.
	\label{eq:f<-ab}
\end{equation}

It follows from Eqs. \eqref{eq:alpha_diff}-\eqref{eq:ab_bound} that
\begin{equation}
	\alpha_E(\vec{r},-\vec{n}) = \beta_E(\vec{r},\vec{n}),
	\label{eq:ab_symmetry}
\end{equation}
and hence
\begin{equation}
	g_E(\vec{r},-\vec{n}) = g_E(\vec{r},\vec{n}), \quad f_E^{\dagger}(\vec{r},-\vec{n}) = f_E(\vec{r},\vec{n}).
	\label{eq:gf_symmetry}
\end{equation}
Another important property is that $\alpha_E$ and $\beta_E$ are monotonically increasing functions of $E$ (when $i\eps$ is not relevant). This follows from the fact that the right-hand sides of Eqs. \eqref{eq:alpha_diff} - \eqref{eq:ab_bound} are monotonous in energy. For $E=0$ we have $\alpha_E = \beta_E = -\pi/2$. Then, for given $\vec{r}$ and $\vec{n}$ a single positive energy $E<\Delta_{\infty}$ may exist, such that $\alpha_E(\vec{r},\vec{n}) + \beta_E(\vec{r},\vec{n}) = 0$. We denote this energy as $\epsilon_g(\vec{r},\vec{n})$. It follows from Eq. \eqref{eq:ab_symmetry} that 
\begin{equation}
	\epsilon_g(\vec{r},-\vec{n}) = \epsilon_g(\vec{r},\vec{n}).
	\label{eq:eg_symmetry}
\end{equation}
Moreover, by adding Eq. \eqref{eq:alpha_diff} to \eqref{eq:beta_diff} one finds that
\begin{equation}
	\hbar v_F \frac{\partial (\alpha_E + \beta_E)}{\partial s} = 4 \Delta(\vec{r}) \sin \! \left( \! \frac{\alpha_E - \beta_E}{2} \! \right) \! \sin \! \left( \! \frac{\alpha_E + \beta_E}{2} \! \right) \!,
	\label{eq:alpha+beta}
\end{equation}
which means that if $\alpha_E + \beta_E = 0$ at some point, then this sum vanishes on a whole classical trajectory passing through this point. As a consequence, $\epsilon_g(\vec{r},\vec{n})$ is constant on each line parallel to $\vec{n}$.

Using Eqs. \eqref{eq:GER_qq}, \eqref{eq:FE_qq}, \eqref{eq:g<-ab} - \eqref{eq:gf_symmetry}, we can write the Green functions with coinciding arguments as
\begin{eqnarray}
	& G_{ER}^{(0)}(\vec{r},\vec{r}) = -\frac{mk_F}{2\pi \hbar^2} \int \! \cot \! \left( \frac{\alpha_E(\vec{r},\vec{n}) + \beta_E(\vec{r},\vec{n})}{2} \right) \! \frac{d^2 \vec{n}}{4\pi}, & \label{eq:GER<-ab} \\
	& F_E^{\dagger (0)} (\vec{r},\vec{r}) =  \frac{imk_F}{2\pi \hbar^2} \int \frac{f_E^{\dagger}(\vec{r},\vec{n}) + f_E^{\dagger}(\vec{r},-\vec{n})}{2} \frac{d^2 \vec{n}}{4\pi} & \nonumber \\
	& = -\frac{mk_F}{2\pi \hbar^2} \int \frac{\cos \left(\frac{\alpha_E(\vec{r},\vec{n}) - \beta_E(\vec{r},\vec{n})}{2} \right)}{\sin \left( \frac{\alpha_E(\vec{r},\vec{n}) + \beta_E(\vec{r},\vec{n})}{2} \right)} \frac{d^2 \vec{n}}{4\pi}.&
	\label{eq:FE<-ab}
\end{eqnarray}
If $\abs{E} < \epsilon_g(\vec{r},\vec{n})$ for all $\vec{n}$, the integrand in Eq. \eqref{eq:GER<-ab} is real and hence $G_{ER}^{(0)}(\vec{r},\vec{r})$ is real. However, if $\abs{E} = \epsilon_g(\vec{r},\vec{n})$ for some $\vec{n}$, then the imaginary term $i\eps$ in Eqs. \eqref{eq:alpha_beta} - \eqref{eq:ab_bound} becomes relevant, and $G_{ER}^{(0)}(\vec{r},\vec{r})$ acquires an imaginary part. From this we conclude that the local gap in the density of states $E_{\mathrm{loc}}(\vec{r})$ (without impurity) is given by
\begin{equation}
	E_{\mathrm{loc}}(\vec{r}) = \min_{\vec{n}} \epsilon_g(\vec{r},\vec{n}).
	\label{eq:E_loc}
\end{equation}
Then, the global gap is
\begin{equation}
	E_g = \min_{\vec{r}} E_{\mathrm{loc}}(\vec{r}).
	\label{eq:Eg_loc}
\end{equation}

Let us denote as $\vec{n}_0(\vec{r})$ a unit vector that satisfies the equation
\begin{equation}
	E_{\mathrm{loc}}(\vec{r}) = \epsilon_g(\vec{r},\vec{n}_0(\vec{r}))
	\label{eq:n_0}
\end{equation}
According to Eq. \eqref{eq:eg_symmetry}, the solutions of Eq. \eqref{eq:n_0} come in pairs of $\vec{n}_0$ and $-\vec{n}_0$. This equation may have more than two solutions, however in the absence of rotational symmetry this is extremely unlikely. For now we will assume that only two vectors $\vec{n}_0$ satisfy Eq. \eqref{eq:n_0}. Then, for $E \approx E_{\mathrm{loc}}(\vec{r})$ the main contributions to the integrals in Eqs. \eqref{eq:GER<-ab} and \eqref{eq:FE<-ab} come from $\vec{n} \approx \pm \vec{n}_0$. For $E \approx \epsilon_g(\vec{r},\vec{n})$ we can make use of the linear in $E - \epsilon_g(\vec{r},\vec{n})$ expansion
\begin{equation}
	\sin \left( \frac{\alpha_E(\vec{r},\vec{n}) + \beta_E(\vec{r},\vec{n})}{2} \right) \approx  A(\vec{r},\vec{n}) [E + i\eps - \epsilon_g(\vec{r},\vec{n})],
	\label{eq:E-eg_linear}
\end{equation}
where $A(\vec{r},\vec{n})$ is a positive function. Then, for $E \approx E_{\mathrm{loc}}(\vec{r})$ Eq. \eqref{eq:GER<-ab} yields
\begin{eqnarray}
	& G_{ER}^{(0)}(\vec{r},\vec{r}) \approx -\frac{mk_F}{8\pi^2 \hbar^2 A(\vec{r},\vec{n}_0(\vec{r}))} & \nonumber \\
	& \times \int\limits_{\epsilon_g(\vec{r},\vec{n}) - E_{\mathrm{loc}}(\vec{r}) < \epsilon_0} \frac{d^2 \vec{n}}{E + i\eps - \epsilon_g(\vec{r},\vec{n})}, &
	\label{eq:GER_n0}
\end{eqnarray}
where $\epsilon_0$ is some energy cut-off such that $\abs{E - E_{\mathrm{loc}}(\vec{r})} \ll \epsilon_0 \ll E_g$. For the evaluation of the integral in Eq. \eqref{eq:GER_n0} let us use an orthonormal coordinate system $xyz$ such that the $z$ axis is directed along $\vec{n}_0(\vec{r})$, and for $\vec{n} \approx \vec{n}_0(\vec{r})$ the following Taylor polynomial approximation holds:
\begin{equation}
	\epsilon_g(\vec{r},\vec{n}) \approx E_{\mathrm{loc}}(\vec{r}) + \frac{1}{2} \frac{\partial^2 \epsilon_g}{\partial n_x^2} n_x^2 + \frac{1}{2} \frac{\partial^2 \epsilon_g}{\partial n_y^2} n_y^2,
	\label{eq:epsilon_Taylor}
\end{equation}
where $\partial^2 \epsilon_g/\partial n_x^2 >0$ and $\partial^2 \epsilon_g/\partial n_y^2 >0$ due to Eq. \eqref{eq:E_loc}. To integrate in Eq. \eqref{eq:GER_n0}, it is convenient to use the integration variables $N$ and $\varphi$, defined via
\begin{equation}
	n_x = N \left( \frac{\partial^2 \epsilon_g}{\partial n_x^2} \right)^{-1/2} \!\!\!\!\!\!\!\!\!\!\!\! \cos \varphi, \quad n_y = N \left( \frac{\partial^2 \epsilon_g}{\partial n_y^2} \right)^{-1/2} \!\!\!\!\!\!\!\!\!\!\!\! \sin \varphi.
	\label{eq:N_and_phi}
\end{equation}
The result of integration is
\begin{eqnarray}
	& G_{ER}^{(0)}(\vec{r},\vec{r}) \approx \frac{mk_F}{2\pi \hbar^2 A(\vec{r},\vec{n}_0(\vec{r})) \sqrt{\det M(\vec{r})}} & \nonumber \\
	& \times \left[ \pi i \Theta(E - E_{\mathrm{loc}}(\vec{r})) + \ln \frac{\epsilon_0}{\abs{E - E_{\mathrm{loc}}(\vec{r})}} \right], &
	\label{eq:GER(E_loc)}
\end{eqnarray}
where $\Theta(x)$ is the Heaviside step function, and
\begin{equation}
	M(\vec{r}) = \left. \left(
	\begin{array}{cc}
	  \frac{\partial^2 \epsilon_g}{\partial n_x^2} & \frac{\partial^2 \epsilon_g}{\partial n_x \partial n_y} \\
		\frac{\partial^2 \epsilon_g}{\partial n_x \partial n_y} & \frac{\partial^2 \epsilon_g}{\partial n_y^2}
	\end{array} \right)
	\right|_{\vec{n} = \vec{n_0}(\vec{r})}.
	\label{eq:M}
\end{equation}
A direct consequence of Eq. \eqref{eq:GER(E_loc)} is the presence of a finite jump of the local density of states at $E = E_{\mathrm{loc}}(\vec{r})$. This observation is consistent with Eqs. \eqref{eq:nu_in} and \eqref{eq:nu_out}.

Similarly to Eq. \eqref{eq:GER(E_loc)}, we may obtain from Eq. \eqref{eq:FE<-ab} that
\begin{equation}
	F_E^{\dagger (0)} (\vec{r},\vec{r}) \approx G_{ER}^{(0)}(\vec{r},\vec{r}) \cos \alpha_E(\vec{r},\vec{n}_0(\vec{r}))
	\label{eq:FE(E_loc)}
\end{equation}
for $E \approx E_{\mathrm{loc}}(\vec{r})$. Because of the logarithmic divergence of $G_{ER}^{(0)}(\vec{r},\vec{r})$ at $E = E_{\mathrm{loc}}(\vec{r})$ one may see that for $\vec{r}_i = \vec{r}$
\begin{equation}
	\lim_{E \to E_g-0} {\cal D}_{\uparrow-}(E) = +\infty,
	\label{eq:2states_+1}
\end{equation}
unless $\alpha_E(\vec{r},\vec{n}_0(\vec{r})) = 0$. Similarly, one can prove that
\begin{equation}
	\lim_{E \to -E_g+0} {\cal D}_{\uparrow+}(E) = -\infty.
	\label{eq:2states_-1}
\end{equation}
Thus, the assumptions that we made concerning the order parameter profile provide the sufficient conditions for the existence of two bound states localized on a point impurity.

The case $\alpha_E(\vec{r},\vec{n}_0(\vec{r})) = 0$ corresponds to a very specific placement of the impurity. This happens, for example, when the position of the impurity is the center of inversion symmetry for $\Delta(\vec{r})$: $\Delta(\vec{r}_i + \vec{R}) = \Delta(\vec{r}_i - \vec{R})$ for any vector $\vec{R}$. In this situation there may be less than two impurity states, and their number depends on the values of the scattering phases.

To conclude, we briefly consider the case when the impurity is placed in the center of spherical symmetry of $\Delta(\vec{r})$, such that $\Delta(\vec{r}_i + \vec{R}_1) = \Delta(\vec{r}_i + \vec{R}_2)$ when $\abs{\vec{R}_1} = \abs{\vec{R}_2}$. Then the integrands in Eqs. \eqref{eq:GER<-ab} and \eqref{eq:FE<-ab} do not depend on $\vec{n}$, and additionally $\beta_E(\vec{r}_i,\vec{n}) = \alpha_E(\vec{r}_i,\vec{n}) \equiv \alpha_E(\vec{r}_i)$. Then
\begin{equation}
	G_{ER}^{(0)}(\vec{r}_i,\vec{r}_i) = -\frac{mk_F}{2\pi \hbar^2} \cot \alpha_E(\vec{r}_i),
	\label{eq:GER_spherical}
\end{equation}
\begin{equation}
	F_E^{\dagger (0)} (\vec{r}_i,\vec{r}_i) = -\frac{mk_F}{2\pi \hbar^2} \sin^{-1} \alpha_E(\vec{r}_i).
	\label{eq:FE_spherical}
\end{equation}
The equation ${\cal D}_{\uparrow}(E) = 0$ then yields
\begin{equation}
	\sin\left( \alpha_E(\vec{r}_i) - \alpha_{\downarrow} + \alpha_{\uparrow} \right) = 0.
	\label{eq:alpha_E}
\end{equation}
The function $\alpha_E$ is a monotonically increasing function of energy, and at $E = \pm E_{\mathrm{loc}}(\vec{r}_i)$ it takes the values
\begin{equation}
	\alpha_{E_{\mathrm{loc}}(\vec{r}_i)}(\vec{r}_i) = 0, \quad \alpha_{-E_{\mathrm{loc}}(\vec{r}_i)}(\vec{r}_i) = -\pi.
	\label{eq:+-E_loc}
\end{equation}
This means that for $E \in (-E_{\mathrm{loc}}(\vec{r}_i),E_{\mathrm{loc}}(\vec{r}_i))$ Eq. \eqref{eq:alpha_E} has one solution when $\alpha_{\uparrow} \neq \alpha_{\downarrow}$ and no solutions when $\alpha_{\uparrow} = \alpha_{\downarrow}$. Thus, a magnetic impurity induces one subgap state, and a nonmagntic impurity does not induce subgap states in this case.

%\makeatletter
%\bibliographystyle{aipnum4-1}
%\bibliography{NSphere}
%\makeatother

%merlin.mbs aipnum4-1.bst 2010-07-25 4.21a (PWD, AO, DPC) hacked
%Control: key (0)
%Control: author (8) initials jnrlst
%Control: editor formatted (1) identically to author
%Control: production of article title (-1) disabled
%Control: page (0) single
%Control: year (1) truncated
%Control: production of eprint (0) enabled
%

\end{document}